RESEARCH ARTICLE

# PyNoetic: A modular python framework for no-code development of EEG brain-computer interfaces

**Gursimran Singh**[1], **Aviral Chharia**[2*], **Rahul Upadhyay**[1*], **Vinay Kumar**[1], **Luca Longo**[3]

**1** Electronics and Communication Engineering Department, Thapar Institute of Engineering and Technology, Patiala, India, **2** Department of Mechanical Engineering, Carnegie Mellon University, Pittsburgh, Pennsylvania, United States of America, **3** Artificial Intelligence and Cognitive Load Lab, School of Computer Science, TU Dublin, Dublin, Ireland

* achharia@andrew.cmu.edu (AC); rahul.upadhyay@thapar.edu (RU)

**Data availability statement:** The source code of PyNoetic, the test dataset, as well as the user documentation, is released to the public at https://github.com/NeuroDiag/PyNoetic-official.

**Funding:** The author(s) received no specific funding for this work.

**Competing interests:** The authors have declared that no competing interests exist.

## Abstract

Electroencephalography (EEG)-based Brain-Computer Interfaces (BCIs) have emerged as a transformative technology with applications spanning robotics, virtual reality, medicine, and rehabilitation. However, existing BCI frameworks face several limitations, including a lack of stage-wise flexibility essential for experimental research, steep learning curves for researchers without programming expertise, elevated costs due to reliance on proprietary software, and a lack of all-inclusive features leading to the use of multiple external tools affecting research outcomes. To address these challenges, we present PyNoetic, a modular BCI framework designed to cater to the diverse needs of BCI research. PyNoetic is one of the very few frameworks in Python that encompasses the entire BCI design pipeline, from stimulus presentation and data acquisition to channel selection, filtering, feature extraction, artifact removal, and finally simulation and visualization. Notably, PyNoetic introduces an intuitive and end-to-end GUI coupled with a unique pick-and-place configurable flowchart for no-code BCI design, making it accessible to researchers with minimal programming experience. For advanced users, it facilitates the seamless integration of custom functionalities and novel algorithms with minimal coding, ensuring adaptability at each design stage. PyNoetic also includes a rich array of analytical tools such as machine learning models, brain-connectivity indices, systematic testing functionalities via simulation, and evaluation methods of novel paradigms. PyNoetic's strengths lie in its versatility for both offline and real-time BCI development, which streamlines the design process, allowing researchers to focus on more intricate aspects of BCI development and thus accelerate their research endeavors.

# 1 Introduction

## 1.1 Background

Brain-computer interfaces (BCIs) translate neural activity into computer-executable commands and create a connection between the human brain and computer devices. BCIs hold immense potential to revolutionize several domains, including virtual reality (VR), robotics, entertainment, medicine, and rehabilitation. For example, patients with neurological disorders, like Cerebral Palsy [1], Amyotrophic Lateral Sclerosis [2], and traumatic brain/spinal cord injuries, who experience a compromised neural pathway governing muscle control, can benefit immensely from BCIs [3,4]. BCIs can offer a potential solution by establishing an alternative communication pathway between the patient's brain and the external world, circumventing the compromised nervous system [5,6]. However, the intricate nature of the human brain, coupled with inherent temporal variations [7], underscores the importance of tailoring BCIs to specific disabilities. A BCI system designed for one disability may not effectively address other disabilities. Even within the same disability cohort, the performance and efficacy of BCI systems exhibit significant variability [8]. Moreover, BCIs developed in controlled lab settings often lack the versatility [9] needed for widespread adoption, highlighting the necessity for rapidly prototyping BCIs uniquely customized to individual users.

The development of Electroencephalogram (EEG)-based non-invasive BCIs is preferred over invasive ones due to their perceived safety. An end-to-end EEG-based BCI system typically consists of *(i)* stimuli presentation, *(ii)* EEG data recording/acquisition, *(iii)* EEG signal processing, which includes channel selection, pre-processing, consequent feature extraction and classification, and the *(iv)* generation of an executable computer command. Fig 1 illustrates a schematic representation of a typical BCI system. Existing open-source

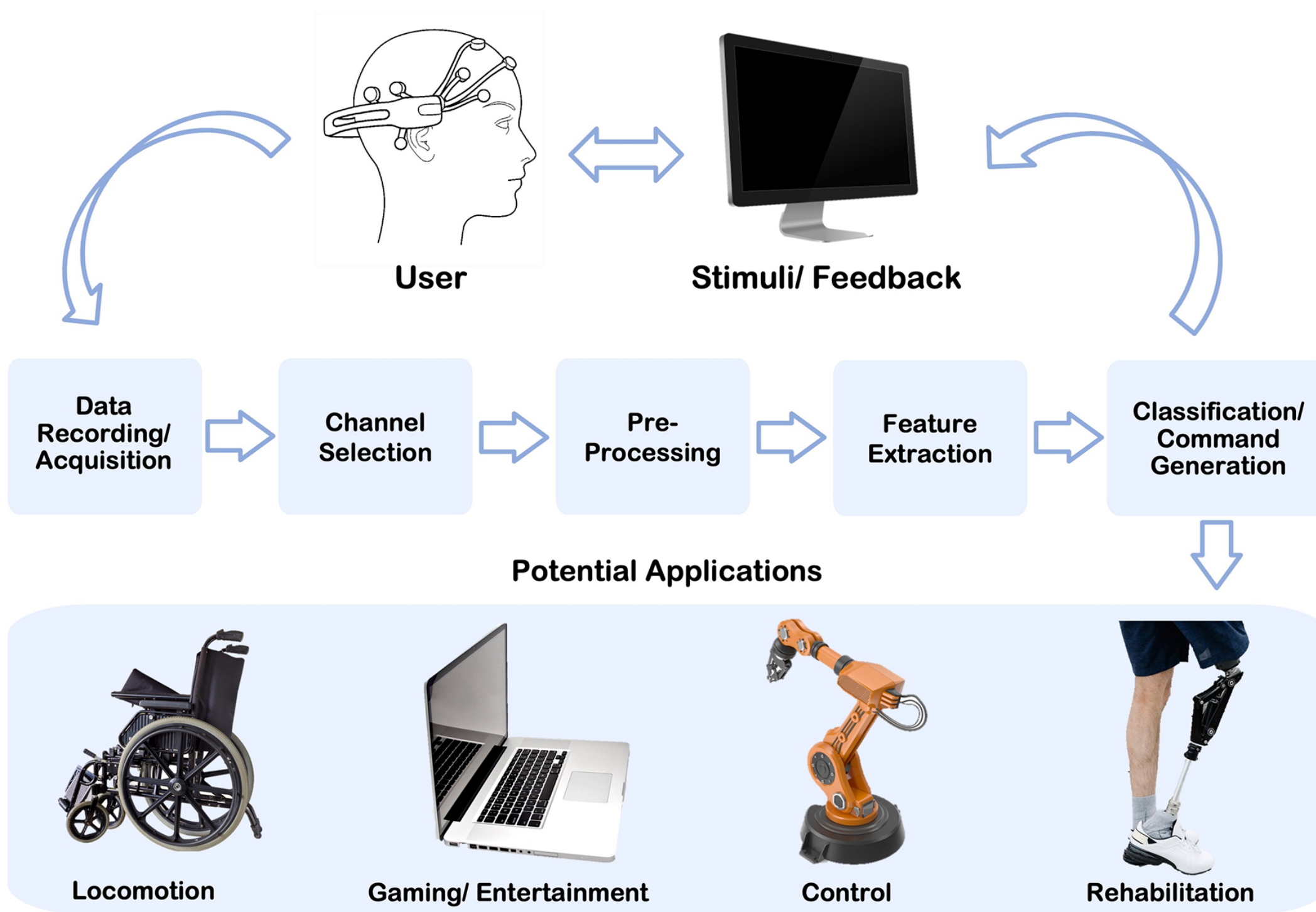

**Fig 1. A typical BCI system depicting the flow and processing of EEG data.**

BCI frameworks exhibit limitations that impede their suitability for comprehensive BCI development.

### 1.2 Challenges and motivation

Testing experimental paradigms often involves coding intricate software (SW) frameworks or toolboxes that lead to a significant barrier in the rapid development of scalable BCI solutions. BCI system development is a multifaceted endeavor [10] requiring expertise across diverse domains, such as neuroscience, biomedical signal processing, embedded systems, ML, and artificial intelligence (AI). A number of BCI frameworks have been proposed including MNE-Python [11], Wyrm Ecosystem [12], BioPyC [13], Gumpy [14], etc. At present, very few BCI frameworks support: *(i)* the development of end-to-end non-invasive EEG-based BCI systems. Further, existing BCI frameworks suffer from critical limitations, including *(ii)* a lack of 'stage-wise' flexibility crucial for experimental research, *(iii)* missing all-inclusive features leading to reliance on multiple external tools, which results in *(iv)* elevated research costs stemming from reliance on multiple proprietary SW, high technological barriers, and may affect research outcomes. *(v)* Lastly, most existing frameworks appeal more to experienced programmers compared to beginners in the field due to the high learning curve in coding intricate SW frameworks. These limitations underscore the ongoing challenges in creating versatile and robust BCI frameworks.

### 1.3 Contribution

To address these challenges, we present PyNoetic, a *novel* open-source, highly modular, customizable, GUI-aided BCI framework built in Python (see Fig 2). PyNoetic primarily aims to streamline the development and prototyping of BCIs [15]. Addressing substantial shortcomings observed in previous frameworks, PyNoetic provides researchers with a stand-alone solution, offering end-to-end capabilities ranging from stimulus presentation and data acquisition to classification and feedback. To the best of our knowledge, PyNoetic offers the following contributions:

1. *End-to-end BCI development*, covering every phase from generating custom stimuli to recording new datasets, implementing channel selection algorithms for complexity reduction, applying filtering and pre-processing techniques for artifact removal, and featuring an extensive feature extraction module. PyNoetic also incorporates popular classifiers and includes a simulation for evaluating the efficacy of the BCI under development.
2. *developed in Python*. Unlike several BCI frameworks primarily developed in languages like C++ (such as xBCI [16], BF+ [17], BCI2000 [18], OpenVibe [19]), PyNoetic stands out as a comprehensive BCI framework in Python. With Python's increasing popularity among BCI researchers, particularly for ML and deep learning, PyNoetic offers significant value. It also provides cross-platform support on Linux, Windows, and macOS, ensuring accessibility to a wide range of researchers.
3. *modular* design, carefully dividing the framework into seven modules based on the expertise of its users, facilitating a plug-and-play approach with effortless SW updates. This enables users to test their BCI systems within a simulated environment, adjust parameters, and observe the effects on test results.
4. *a GUI* allowing users to create and design a "real-time" BCI by simply interacting with it and modifying parameters, unlike existing frameworks. It also includes functionality for *programmable flowchart in online mode*, implementing a pick-and-place architecture, where users can select various stages from the EEG data flow and integrate them into their pipeline.
5. *includes integrated data recording*, providing the capability to present stimuli and dynamically adjust parameters during data acquisition. Recorded data is automatically segmented according to user-defined parameters, simplifying the creation of new datasets for future research.
6. *functionality suitable for both programming novices and experts*. Featuring an intuitive GUI with distinct tabs representing different stages in the BCI development, enabling parameter tuning through simple mouse interactions, and requiring no prior programming expertise.
7. *platform support*. PyNoetic provides cross-platform support on Linux, Windows, and macOS, ensuring accessibility to a wide range of researchers. PyNoetic has been tested on Ubuntu 24.04 LTS, Windows 11, and macOS Sequoia. It should be noted that older versions of these systems might not be supported by some EEG acquisition devices due to the lack of hardware drivers on that platform.

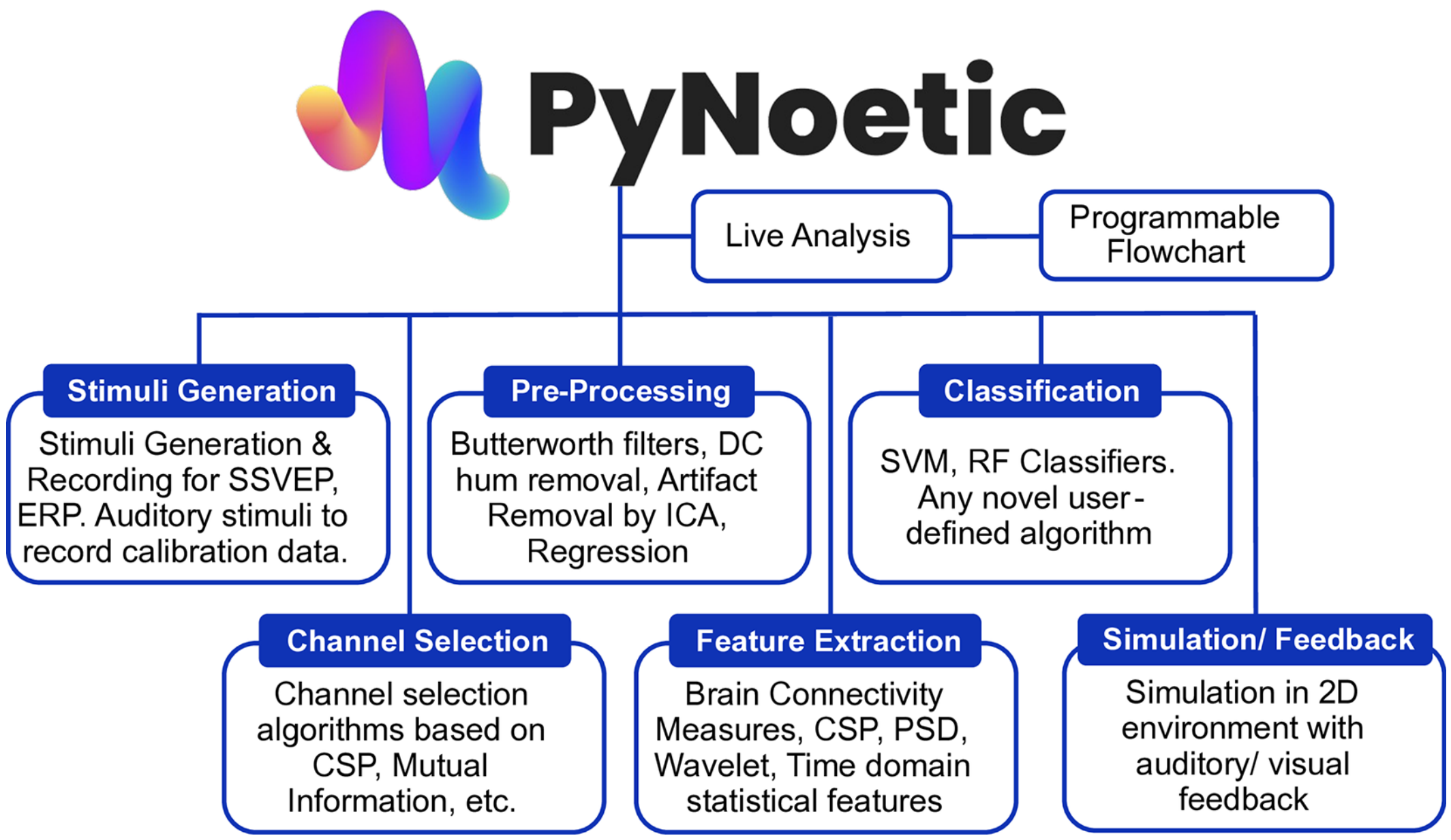

**Fig 2. Overview of functionality supported by PyNoetic and its various modules, including the live analysis and programmable flowchart.**

The remainder of this article is as follows: Sect 2 presents the related works, and Sect 3 introduces the overall design of the PyNoetic, structured to reflect the sequential information flow in the development of a BCI system. This section also delves into the GUI and other essential functionalities integrated within PyNoetic, providing insights into the rationale behind their inclusion. Sect 4 details the diverse SW design constructs utilized. Sect 6 presents the experimental verification and case studies, while Sect 7 offers a comprehensive discussion of the framework, comparing it with existing frameworks, highlighting its strengths, and critically examining the design principles that underpin its development. Lastly, Sect 8 presents the conclusion, and Sect 9 recognizes the limitations and outlines avenues for future research.

| | List of Abbreviations |
|---|---|
| **Abbreviation** | **Full name** |
| BSS | Blind Source Separation |
| CNN(s) | Convolutional Neural Network(s) |
| CSP | Common Spatial Patterns |
| DFA | Detrended Fluctuation Analysis |
| DT | Decision Tree |
| DWT | Discrete Wavelet Transform |
| ERP(s) | Event-Related Potential(s) |
| FD | Fractal Dimensions |
| FFT | Fast Fourier Transform |
| IC(s) | Independent Component(s) |
| ICA | Independent Component Analysis |
| ITR | Information Transfer Rate |
| LSTM | Long Short-Term Memory |
| MI | Motor Imagery |
| NB | Naive Bayes |
| PLV | Phase Locking Value |
| PSD | Power Spectral Density |
| RF | Random Forest |
| SSVEP | Steady-State Visual Evoked Potential |
| STFT | Short-Time Fourier Transform |
| SVM | Support Vector Machine |
| TDF(s) | Time Domain Feature(s) |

## 2 Related works

In this section, we discuss popular BCI frameworks along with their supported functionality and features. MNE-Python [11] is an open-source Python library that handles human neuro-physiological data, primarily focusing on denoising it, applying filters, and calculating connectivity estimates. It is scripting-based and has the ability to visualize processing stages, but it offers limited GUI support. It uses popular Python libraries like Matplotlib [20] and SciPy [21] to implement various signal-processing algorithms.

BCI2000 [18] is a popular C++ software framework that is structured around a flexible module-based system that can represent any BCI system by leveraging its extensive scripting capabilities and graphical user interface. OpenVibe [19] is another C++ BCI platform that supports modular architecture along with a programmable flowchart, making it adoptable by a diverse group of users. While both OpenVibe and BCI2000 are implemented in C++, they enable support for Python and other programming languages through interfaces and plugins for prototyping.

Wyrm Ecosystem [12] provides features for processing and plotting data. Wyrm, along with Mushu [22], a signal acquisition library, and PyFF [23], a framework for stimuli and feedback presentation, provides a Python environment for displaying stimuli, analyzing acquired signals, and providing real-time feedback to the user. Wyrm makes designing experimental paradigms programmable and supports connections to external hardware such as eye trackers. It is one of the very few software programs that support microstate analysis of EEG. Mushu supports a variety of hardware and outputs EEG data in the form of Numpy arrays [24], inherently supported by PyNoetic.

BioPyC [13] is another popular Python framework for offline analysis of biomedical signals. It has a GUI based on Jupyter (.ipynb) Notebook and can load, process, and classify datasets, and visualize the obtained results. BioPyC [13] is entirely GUI-based and requires no prior programming experience. Each step of the GUI has instructions to guide the users, thus making the software extremely beginner-friendly. BioPyc [13] offers another notable feature through its integration with established bioinformatics tools, enabling their seamless incorporation into the workflow.

Gumpy [14] is a framework for analyzing EEG and EMG signals that can plot, process, and classify data. Gumpy has extensive support for deep learning models such as Convolutional Neural Networks (CNN), Long Short-Term Memory (LSTM), etc. Gumpy also incorporates experimental recording paradigms such as SSVEP and motor imagery to record new datasets for Hybrid BCI. Gumpy has been demonstrated to be competitive in offline analysis as well as real-time applications. In addition to feature extraction and classification, it includes advanced features such as Riemannian distance and phase locking value (PLV).

Medusa [25] offers a Python-based ecosystem equipped with a wide range of ready-to-use BCI paradigms, such as P-300 spellers, motor imagery, and biofeedback, along with several signal processing functions. Medusa also enables sharing user-defined custom experiments on their website to promote reproducibility.

HappyFeat [26] aims to be a software assistant, helping optimize BCI pipelines by extracting and selecting classification features. HappyFeat leverages OpenVibe to achieve the complete BCI workflow from signal acquisition to online classification. Support for other signal processing platforms, such as Timeflux [27], a Python package for real-time signal processing, is being worked on. NeuXus [28] is another Python package similar to Timeflux for developing real-time pipelines using a configurable nodal architecture. Other Python frameworks of note are BciPy [29] and BCI-HIL [30]. BciPy focuses on ERP-based spelling interfaces for text dictation, leveraging PsychoPy for stimulus presentation, but it can also be used for other paradigms. BCI-HIL is a hardware-independent platform built on the Timeflux package that leverages cloud computing for human-in-loop model training, real-time stimulus control, and transfer learning.

## 3 PyNoetic: Proposed design, features and functionality

### 3.1 Overview

**SW Design.** The portmanteau "PyNoetic" combines "Python," the programming language, with "Noetic," which pertains to the study of the mind and intellect. PyNoetic is a comprehensive framework consisting of seven modules: *(i)* Stimuli Generation and Recording, *(ii)* Channel Selection, *(iii)* Pre-processing, *(iv)* Feature Extraction, *(v)* Classification, *(vi)* Visualization, and *(vii)* Subject Training and Feedback. These modules facilitate both online and offline BCI data analysis.

**Choice of programming language.** Predominantly, the development of BCI frameworks relied on languages such as MATLAB and C++ [31]. While MATLAB offers convenient signal processing tools, its use necessitates a costly license. On the other hand, C++, though open-source, demands advanced programming skills and appeals only to BCI researchers with programming experience. Recently, Python has emerged as an alternative due to its intuitive and easy-to-learn semantics. Additionally, Python supports a vast library ecosystem, which surpasses MATLAB in functionality [20,21,24,32], while being more accessible to novice programmers. PyNoetic is developed in Python (see Fig 3).

**Open distribution.** PyNoetic is released as an open-source framework distributed under the GNU General Public License (GPL) and freely available for non-commercial use. This

```
def train_model(self):
    classifier = Classification(self.channels)
    self.lineEdit_6.clear()
    model_index = self.comboBox_6.currentIndex()
    if model_index == 0:
        acc = classifier.dtree_model()
    self.lineEdit_6.setText(acc)

def dtree_model(self):
    estimator = DecisionTreeClassifier()
    X = self.data[:-1]
    Y = self.data[-1]
    # Set random seed for reproducibility
    np.random.seed(38)
```

(a) Adding a Novel/ Custom ML model to PyNoetic

```
from PyNoetic.live_mode.feature import Classification,
    Preprocess, Feature

class ClassificationNode(CtrlNode):
    nodeName = "Classification"
    algorithm_options = ['Decision Tree', 'Random Forest',
                         'SVM', 'Naive Bayes']
    uiTemplate = [('Algorithm', 'combo',
                   {'values':algorithm_options})]

    def __init__(self, name):
    # Define IO terminals available on this node
        terminals = { 'CIn': dict(io='in'),
                      'COut': dict(io='out')}
    def process(self, CIn, display=True):
        classifier = self.ctrls['Algorithm'].currentText()
        result = Classification.loaded_model.predict(
            np.array(Cin).reshape(1,-1))
     return {'Cout': result}
```

(b) Node for real-time classification using a pre-trained model

```
from PyNoetic.toolbox import Toolbox
from PyNoetic.plot import PlotEEG
def random_gen(self, n, classes): ...
def stimuli(self,period,buffer,cross_period):
    # Random_gen generates a pseudo random stimuli
    df = pd.read_csv("sequence.csv")
    markers = df["Sequence"].to_numpy()
    self.display_cross(cross_period)
    time.sleep(buffer)
    for i in markers:
        if i==0:
            Toolbox.display_left(period)
            time.sleep(buffer)
        elif i==1:
            Toolbox.display_right(period)
            time.sleep(buffer)
        elif i==2:
            Toolbox.display_top(period)
            time.sleep(buffer)
        else:
            Toolbox.display_bottom(period)
            time.sleep(buffer)
# Custom code can be added to modify stimuli
def display_left(self, time): ...
def display_right(self, time): ...
def display_top(self, time): ...
def display_bottom(self, time): ...
```

(c) Stimuli Generation in PyNoetic

```
class Worker(QObject):
    finished = pyqtSignal()
    def start(self):
        os.system("python"+os.path.dirname(os.getcwd()) +
                  "/cortex_api/record.py")
        #Alternative for Lab streaming layer
        """
        streams = pylsl.resolve_stream('type', 'EEG')
        inlet = pylsl.StreamInlet(streams[0])
        """
        self.finished.emit()
```

(d) Parallel thread for recording data using a custom API or Lab Streaming Layer

```
from PyNoetic.live_mode.feature import Classification,
    Preprocess, Feature

class PreProcessingNode(CtrlNode):
    nodeName = "Pre-processing"
    filter_options = ["lowpass","highpass","bandpass","bandstop"]
    uiTemplate = [
        ('Amplitide Threshold', 'spin',
                 {'value': 1.0, 'step': 1.0, 'bounds': [1.0, None]}),
        ('Filter', 'combo', {'values': filter_options}),
        ('Order', 'spin', {'value': 1.0, 'step': 1.0, 'bounds': [1.0, None]}),
        ('Sampling Frequency', 'spin',
                 {'value': 128.0, 'step': 128.0, 'bounds': [128.0, None]}),
        ('Upper CutOff Frequency', 'spin',
                 {'value': 0.0, 'step': 5.0, 'bounds': [0.0, None]}),
        ('Lower CutOff Frequency', 'spin',
                 {'value': 0.0, 'step': 5.0, 'bounds': [0.0, None]})]

    def __init__(self, name):
        # Define the input / output terminals available on this node
        terminals = {'PPIn': dict(io='in'), 'PPOut': dict(io='out')}
        CtrlNode.__init__(self, name, terminals=terminals)

    def process(self, PPIn, display=True):
        proc = Preprocess(checklist,          PPIn)
        filter_choice = str(self.ctrls['Filter'].currentText())
        order =          self.ctrls['Order'].value()
        fs =             self.ctrls['Sampling Frequency'].value()
        freq             = [self.ctrls['Upper CutOff Frequency'].value(),
                            self.ctrls['Upper CutOff Frequency'].value()]
        preprocessed_data = proc.apply_filter(filter_choice, order, fs, freq)
        return {'PPOut': preprocessed_data}
```

(e) PyNoetic Real Time Processing Node

```
library.addNodeType(ChannelSelectionNode, [('Channel Selection',)])
library.addNodeType(PreProcessingNode, [('Pre Processing',)])
library.addNodeType(FeatureExtractionNode, [('Feature Extraction',)])
library.addNodeType(ClassificationNode, [('Classification',)])
```

(f) Adding custom nodes to PyNoetic Flowchart

**Fig 3. Code snippets to add user-specific custom functionality in PyNoetic.** It illustrates the simplicity due to its modular approach, choice of programming language, and other advantages discussed.

licensing choice is driven by the desire to foster a collaborative environment among the BCI community to continuously enhance the framework. The primary objective is to keep the framework updated with the latest state-of-the-art methods, positioning it at the forefront of BCI research. This collaborative approach ensures that PyNoetic remains a dynamic resource for the broader BCI community.

## 3.2 Encapsulating a stage-wise modular design

**(i) PyNoetic's stimuli generation and recording module.** The design of experimental paradigms, particularly for stimulus presentation, is a crucial task in BCI development. Notably, PyNoetic distinguishes itself as one of the very few open-source frameworks that seamlessly unifies both stimuli generation and recording into a single framework (see Fig 4), absent in other popular frameworks like Wyrm [12], BioPyC [13], PyEEG [33], eeglib [34], EEGraph [35] and NeuroKit2 [36]. This seamless integration provides: *(i)* a significant advantage by allowing the utilization of temporal windows, which simplifies the subsequent feature extraction processes. *(ii)* Moreover, it facilitates the automatic division of recorded signals into epochs based on user-defined parameters. *(iii)* An important standalone use case of this module is to generate specific application-tailored training datasets, which is crucial given the non-linear nature of the brain and the multitude of external/internal factors affecting recorded data [37].

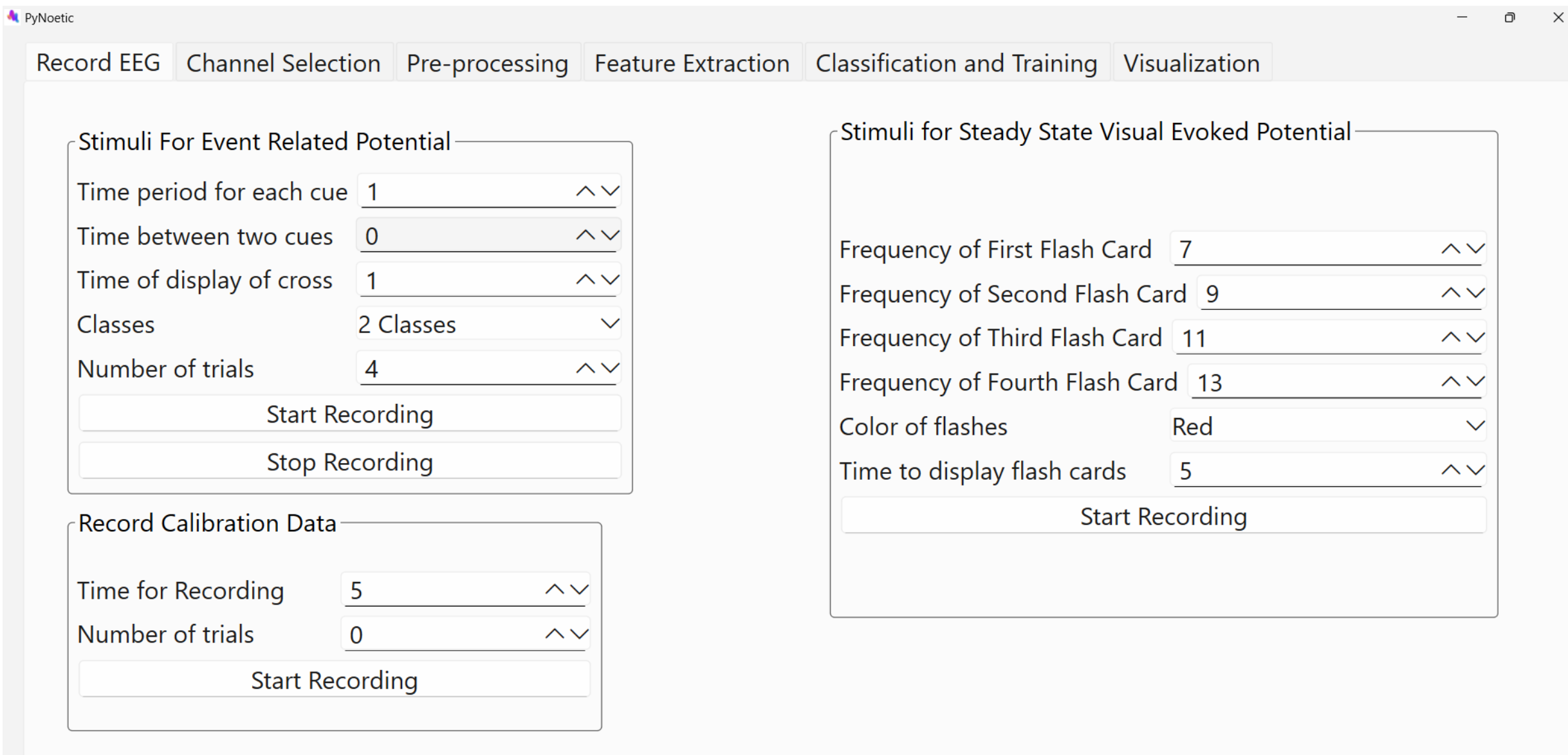

**Fig 4. PyNoetic's stimuli generation and recording module, which supports both ERP and SSVEP.**

PyNoetic's stimuli generation module enables the generation of stimuli for two classes of neural mechanisms: Event-Related Potential (ERP) and Steady-State Visual Evoked Potential (SSVEP). PyNoetic's GUI allows for the real-time modification of the parameters of the generated stimuli. A separate sub-module for calibration data recording is also included. In addition to visual cues, the module supports auditory stimuli.

- *Event-Related Potential (ERP) sub-module:* ERPs are EEG fluctuations that occur in response to visual, auditory, or somatosensory stimuli and are time-locked to the given stimulation [38]. In PyNoetic, the ERP sub-module presents the stimuli on the screen for a defined duration of time. Users can specify the type of stimulus, its duration, inter-stimulus interval, and various other parameters. Table 1 describes the list of tunable ERP parameters. Additionally, any custom image dataset can be utilized to construct visual stimuli. The total experiment duration is calculated using Eq 1:

$$\text{Time} = (\text{Cue time} + \text{Buffer time}) \times \text{Trials} + \text{Fixation time} \tag{1}$$

**Table 1. Tunable parameters in stimuli generation module.** [a]**The fixation cross is displayed once before an experiment starts to help subjects concentrate.** [b]**Different stimuli are displayed equally. If desired by the user, some stimuli can be displayed more/less frequently by assigning them a higher/lower weight.**

| Parameter | Description |
|---|---|
| Cue Time | Duration for which individual stimuli remain visible |
| Buffer Time | Time between two successive stimuli |
| Fixation Time | Time period of fixation cross[a] display |
| Classes | Number of different types of stimuli |
| Trial Count | Total number of stimuli to be flashed in an experiment |
| Weights | Weights for individual stimuli[b] |

- *Steady State visual evoked potential (SSVEP) sub-module:* SSVEP is a widely used paradigm in BCI development. Its underlying principle lies in the presence of the fundamental frequency of visual stimuli being flashed on the screen while recording EEG. In such experiments, harmonics of the fundamental frequency may also be present in the recorded EEG activity [39]. In PyNoetic's stimuli generation module, multiple stimuli are displayed simultaneously on the screen, provided by the user as images. The SSVEP sub-module enables the user to set a frequency for each stimulus displayed on the screen, as well as the total experiment duration. It is advisable to use lower frequencies, as lower frequency evoked responses tend to have higher amplitudes and wider distribution over the head compared to higher frequencies [40].
- *Calibration Data Recording sub-module:* PyNoetic's calibration data recording sub-module is designed to record EEG data, which is subsequently utilized in the framework's preprocessing module to eliminate unwanted artifacts from the recording. A common application of this feature is to filter ocular artifacts from the EEG data. During calibration, the sub-module generates a series of beeps at regular intervals and prompts the user to blink upon hearing a beep. Auditory feedback is preferred over visual feedback in this sub-module, as auditory stimuli tend to elicit faster reactions, thereby reducing latency [41]. Additionally, auditory stimuli help avoid other evoked responses in the recorded EEG signals.

**(ii) PyNoetic's channel selection module.** Achieving real-time performance holds paramount significance in BCI. This necessitates a balance between computational efficiency and functional efficacy. For this, effective channel selection plays a crucial role in subsequent task classification. Removing redundant and noisy EEG channels not only prevents overfitting but also reduces the setup time for experiments, as individual electrodes often require adhesive attachment to the subject's scalp [42]. Moreover, given that the most popular EEG acquisition systems on the market are wireless headsets, reducing the number of active channels can result in decreased system power consumption. This enables researchers to conduct prolonged experiments without interruption. PyNoetic's channel selection module (see Fig 5) employs various criteria that allow users to judiciously select the top '*n*' EEG channels from the available list in the recording for further analysis. These criteria include:

- *Correlation Criteria:* Using the correlation criteria, PyNoetic identifies the top '*n*' channels with the highest correlation. This method assesses the linear dependency between the output and various variables by utilizing Pearson's correlation coefficient (see Eq 2) [43]. Here, $x_i$ represents the $i^{th}$ variable and $Y$ denotes the target output class.

$$R(i) = \frac{cov(x_i, Y)}{sqrt(var(x_i) \times var(Y))} \tag{2}$$

- *Mutual Information:* Mutual information [44] helps uncover dependencies between two variables through their marginal and conditional entropy (see Eq 3). Here, $I$ represents the mutual information, $H(Y)$ denotes the marginal entropy, and $H(Y/X)$ signifies the conditional entropy. The mutual information $I$ equals zero if $X$ and $Y$ are independent of each other, and it surpasses zero if they are dependent.

$$I(X, Y) = H(Y) - H(Y/X) \tag{3}$$

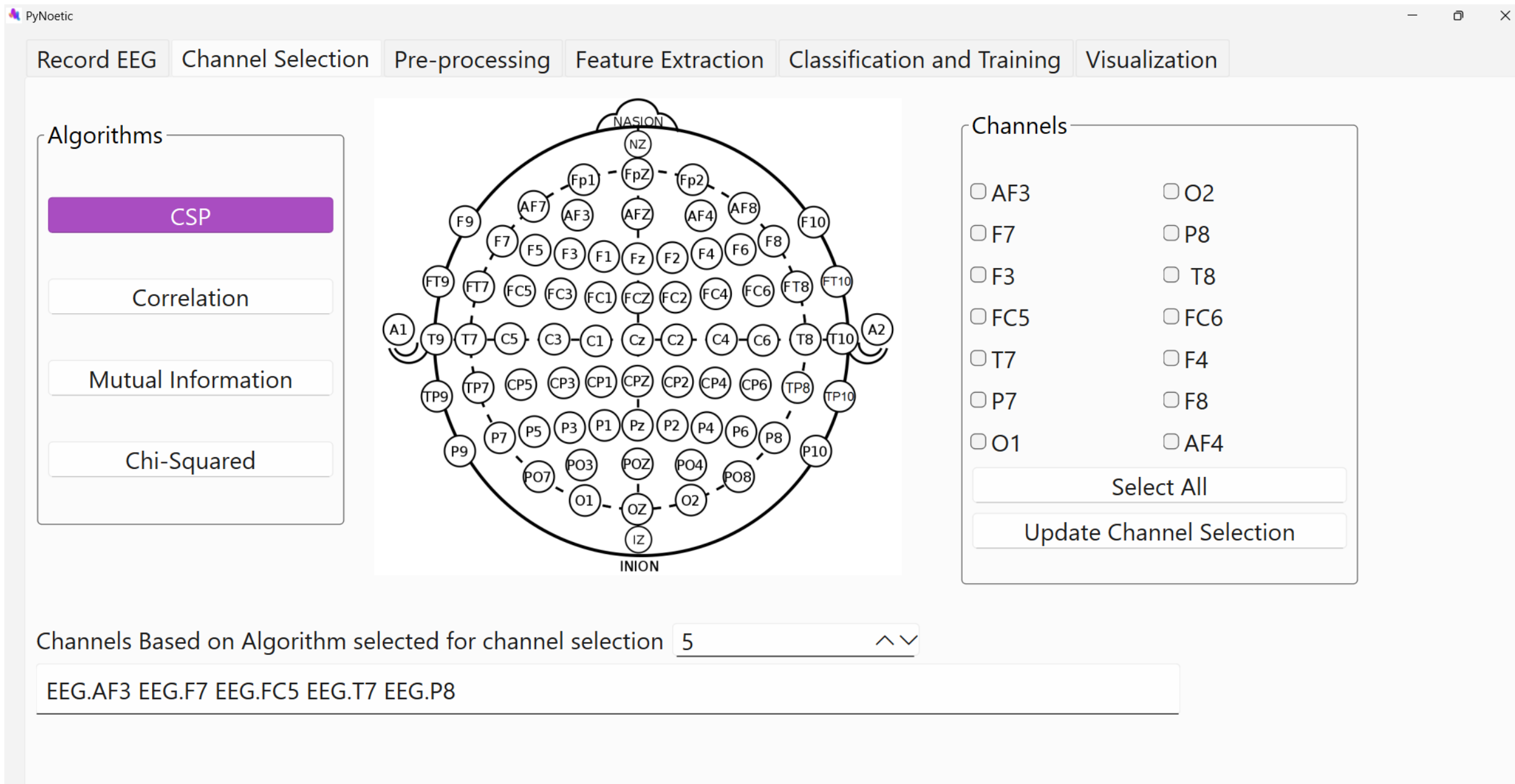

**Fig 5. PyNoetic's channel selection module, which supports various channel selection criteria including CSP, correlation, Mutual Information, and Chi-squared.**

https://doi.org/10.1371/journal.pone.0327791.g005

- *Chi-Squared:* The value of the Chi-Squared test for an EEG channel (see Eq 4) is directly proportional to the channel's relevance for the target class [45]. In this equation, $O_i$ represents the observed value, and $E_i$ denotes the expected value of the variable.

$$X^2 = \sum_{i=1}^{k} \frac{(O_i - E_i)^2}{E_i} \tag{4}$$

- *Common Spatial Patterns (CSP)* PyNoetic also facilitates channel selection based on Common Spatial Patterns (CSP). The spatial patterns derived from the CSP method [46] can be viewed as EEG source distribution vectors. These vectors' peaks are utilized to select the optimal '*n*' channels with the highest correlation for a specific task [47].

PyNoetic provides both offline and online operation capabilities. Therefore, users can utilize the offline mode to identify the most significant EEG channels and subsequently employ only the selected channels during real-time operation.

**(iii) PyNoetic's pre-processing module.** EEG signals are inherently susceptible to various sources of noise, necessitating mitigation to improve the signal-to-noise ratio (SNR) and facilitate accurate EEG feature extraction [48]. Artifactual activity in EEG data can be broadly categorized into physiological and extra-physiological sources, with the latter originating from environmental factors that can be controlled in laboratory settings [49]. However, real-world applications often face challenges due to artifactual components, underscoring the significance of robust EEG artifact removal methods in such contexts [50]. These artifacts manifest as unwanted signals picked up by EEG recording equipment, arising from muscle activity, eye movements, blinks, and other external sources. Removing them is essential to ensure accurate analysis of the underlying neural activity, given that artifacts often exhibit significantly higher amplitudes than ongoing cerebral activity. The pre-processing module of PyNoetic (see Fig 6) encompasses:

- *Filtering:* Filtering integrates digital Butterworth filters for signal filtering, offering users flexibility in configuring various filtering settings. Users can specify the filter type, order, and transition bandwidth in the filter's frequency response. Notably, increasing the filter order enhances filtering precision but also escalates computational demands. To optimize performance, users have the option to specify the sampling frequency of the signal acquisition system, critical for adhering to Nyquist criteria. The filtering module leverages SciPy for implementation [21].
- *Artifact removal:* A common approach for artifact removal and enhancing SNR involves discarding data epochs that surpass a pre-defined amplitude threshold. However, this method proves less effective when dealing with limited data epochs or frequent artifacts [51]. In PyNoetic, two primary techniques are available for the exclusion of noisy epochs:

  ***1. Regression method.*** It involves utilizing calibration data gathered prior to the main experiment to derive regression coefficients for EEG data [52]. These coefficients, computed using NumPy through ordinary least squares, enable the removal of artifacts such as ocular movements while preserving cerebral activity inadvertently recorded during calibration. This method is advantageous for single-channel recordings due to its simplicity compared to techniques like Independent Component Analysis (ICA) [52].

  ***2. Blind source separation method.*** This method decomposes EEG data into maximally Independent Components (ICs), aiming to separate artifact-affected from artifact-free components. Subsequently, artifact-affected and artifact-free ICs are identified, and artifact-free components are combined to yield artifact-free EEG data using the inverse ICA technique [53]. While computationally intensive compared to regression, this method excels in scenarios where calibration data is unavailable. PyNoetic employs MNE's IC-label to identify and discard artifactual ICs, utilizing MNE-ICALabel for automated IC classification using neural networks [53].

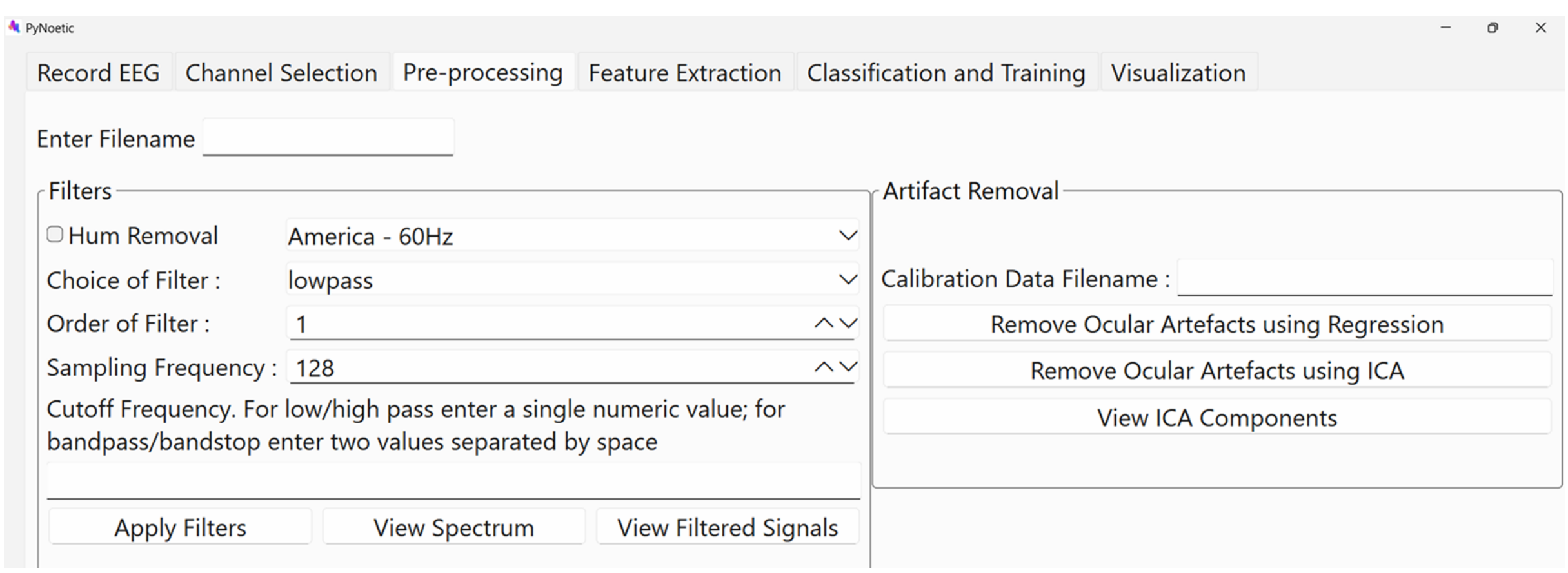

**Fig 6. PyNoetic's pre-processing module, which supports filtering and artifact removal, including ICA.**

https://doi.org/10.1371/journal.pone.0327791.g006

In summary, PyNoetic's pre-processing module integrates advanced techniques for EEG artifact removal, addressing challenges posed by various noise sources to enhance the reliability and accuracy of EEG data analysis in both laboratory and real-world settings.

**(iv) PyNoetic's feature extraction module.** EEG Feature extraction helps in dimensionality reduction, effectively transforming raw EEG data into feature vectors that offer enhanced representations, simplifying subsequent classification tasks. PyNoetic's feature extraction module (refer to Fig 7) encompasses a comprehensive suite of techniques designed to estimate diverse EEG features across time, frequency, time-frequency, and spatial domains. These features and connectivity measures collectively enable researchers to conduct thorough and nuanced analyses of EEG data.

- *Time Domain Features:* In the time domain, PyNoetic supports a wide array of features, including statistical moments such as Mean, Variance, Skewness, and Kurtosis. Additionally, it incorporates non-linear measures such as Fractal Dimensions (Higuchi and Katz), Entropy measures (Shannon entropy, approximate entropy, and sample entropy), Hjorth Parameters, and Detrended Fluctuation Analysis. These measures provide insights into the temporal dynamics and complexity of EEG signals.
- *Frequency Domain Features:* PyNoetic includes essential frequency domain features such as Power Spectral Density, Band Power, and Relative Band Power. These features characterize the spectral content of EEG signals, offering insights into frequency-specific neuronal activities.
- *Time-Frequency Domain Features:* Utilizing techniques like Short-Time Fourier Transform (STFT) and Discrete Wavelet Transform (DWT), PyNoetic computes time-frequency domain features. These features capture and reveal transient and localized changes in brain activity.

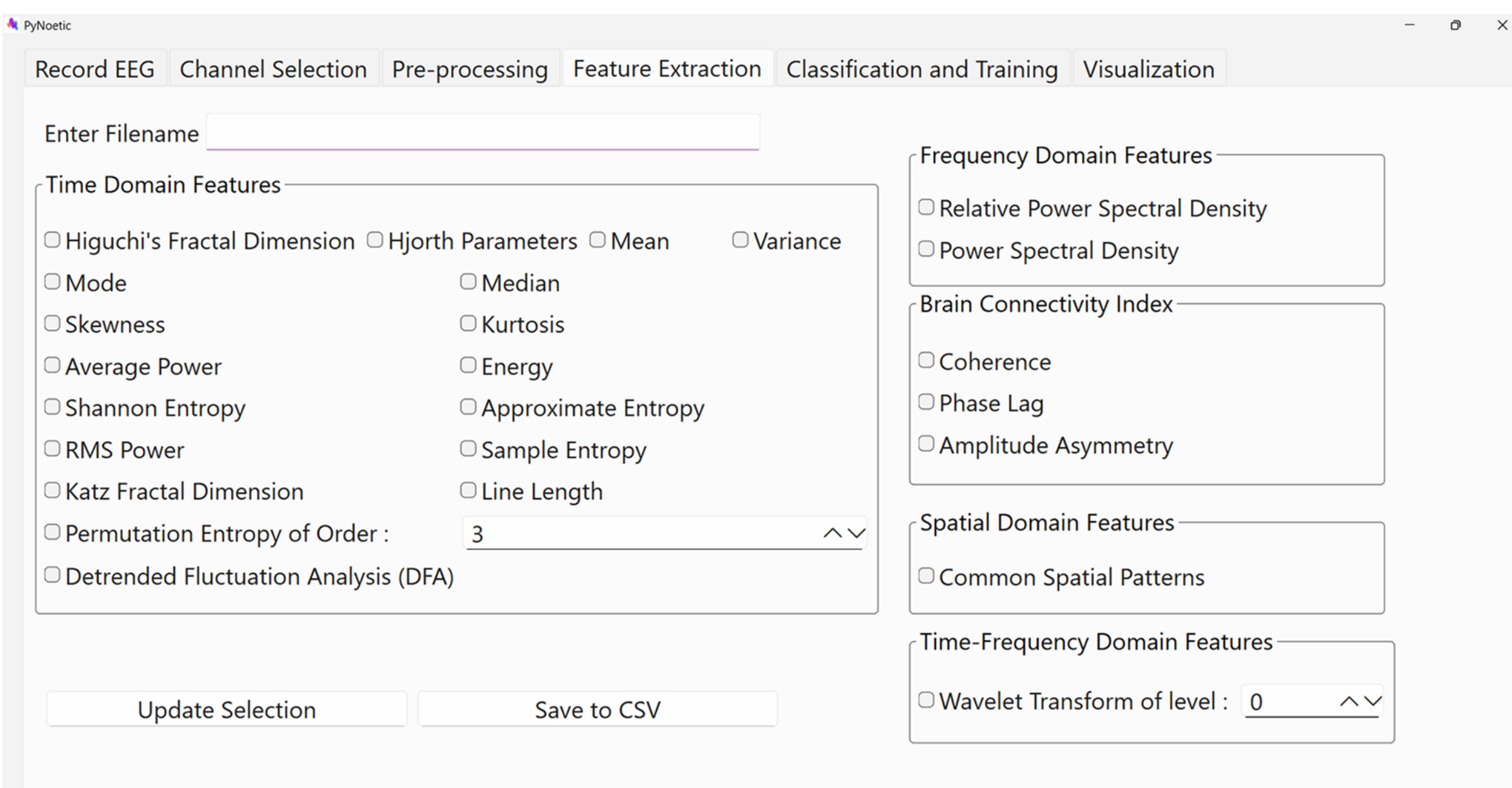

**Fig 7. PyNoetic's feature extraction module that supports time domain features, frequency domain features, time-frequency domain features, spatial features as well as Brain Connectivity measures.**

- *Spatial Domain Features:* Addressing the challenge of limited spatial resolution inherent in EEG due to volume conduction effects [54], which presents a blurred neuronal activity image, PyNoetic employs Common Spatial Patterns (CSP). CSP optimizes spatial filters to maximize the variance of EEG signals across channels, facilitating improved localization of brain activity. This approach is particularly valuable in multi-channel EEG setups for enhancing spatial discrimination and interpretation.
- *Brain Connectivity Measures:* PyNoetic supports important metrics for assessing brain connectivity, including Cross-correlation, Coherence, and Phase Slope Index. They help to quantify the functional relationships and synchronization patterns between different brain regions, providing insights into neural network dynamics and information flow.

**(v) PyNoetic's classification module.** The primary objective of a BCI system is to translate the acquired EEG signals into actionable commands, necessitating the extraction of pertinent features from EEG data and subsequent classification to identify physiological patterns for functional command translation. Various methodologies, including regression and classification, are employed for this purpose, with classification techniques being predominantly favored [55]. PyNoetic provides a suite of ML-based classification models such as Decision Tree (DT), Random Forest (RF), Support Vector Machines (SVM), and Naive Bayes (NB), Riemannian minimum distance to mean (RMDM), alongside deep learning models including EEG-Net, Shallow-Net, and Deep-Net. Covariance matrices lie on a Riemannian manifold and need to be projected on a tangent space to be used as features by classifiers other than RMDM [56]. The flexible architecture of PyNoetic allows effortless integration of novel ML or DL-based classifiers into the framework's core functionality and GUI, **via a single line of function call** (refer to Fig 8). This adaptability facilitates the continuous enhancement of BCI systems tailored to specific user requirements.

**(vi) PyNoetic's simulation module.** The simulation module of PyNoetic (see Fig 9) serves as a critical tool for evaluating the developed pipeline within a 2D simulated environment. Built on the PyGame framework, originally crafted for video game development [57], this module enables comprehensive testing of a two-class Brain-Computer Interface (BCI) system. The simulation scenario closely resembles an obstacle avoidance game where obstacles approach from either side of the screen. The simulation unfolds in three key phases: firstly, it announces the class of the approaching obstacle; secondly, it executes actions based on the BCI system's output, assessing the subject's ability to avoid the obstacle; and finally, it provides auditory and visual feedback to enhance user engagement. Moreover, PyNoetic empowers users to customize simulation parameters to suit specific experimental requirements, including variables such as inter-obstacle intervals, obstacle quantity, and the sequence of on-screen obstacle presentation. This flexibility ensures adaptability for diverse research and development needs in BCI technology.

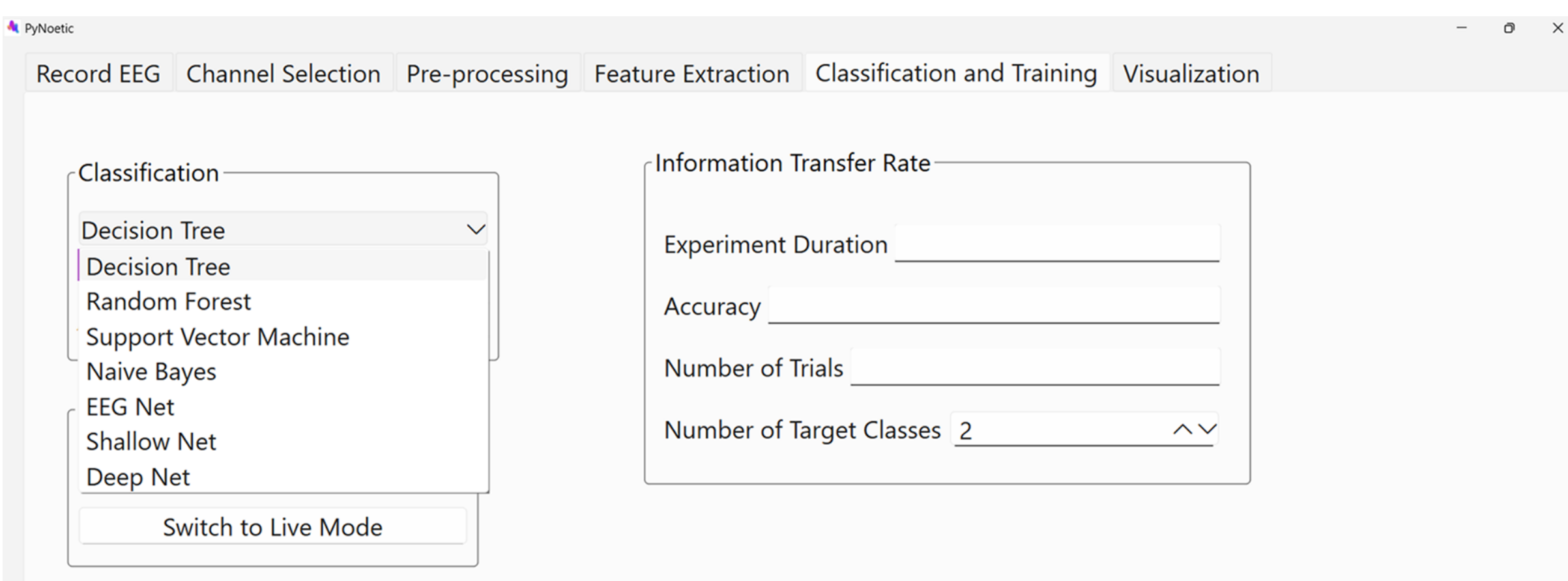

**Fig 8. PyNoetic's classification module supports a range of popular and widely used ML classification models with just a single line of function call.**

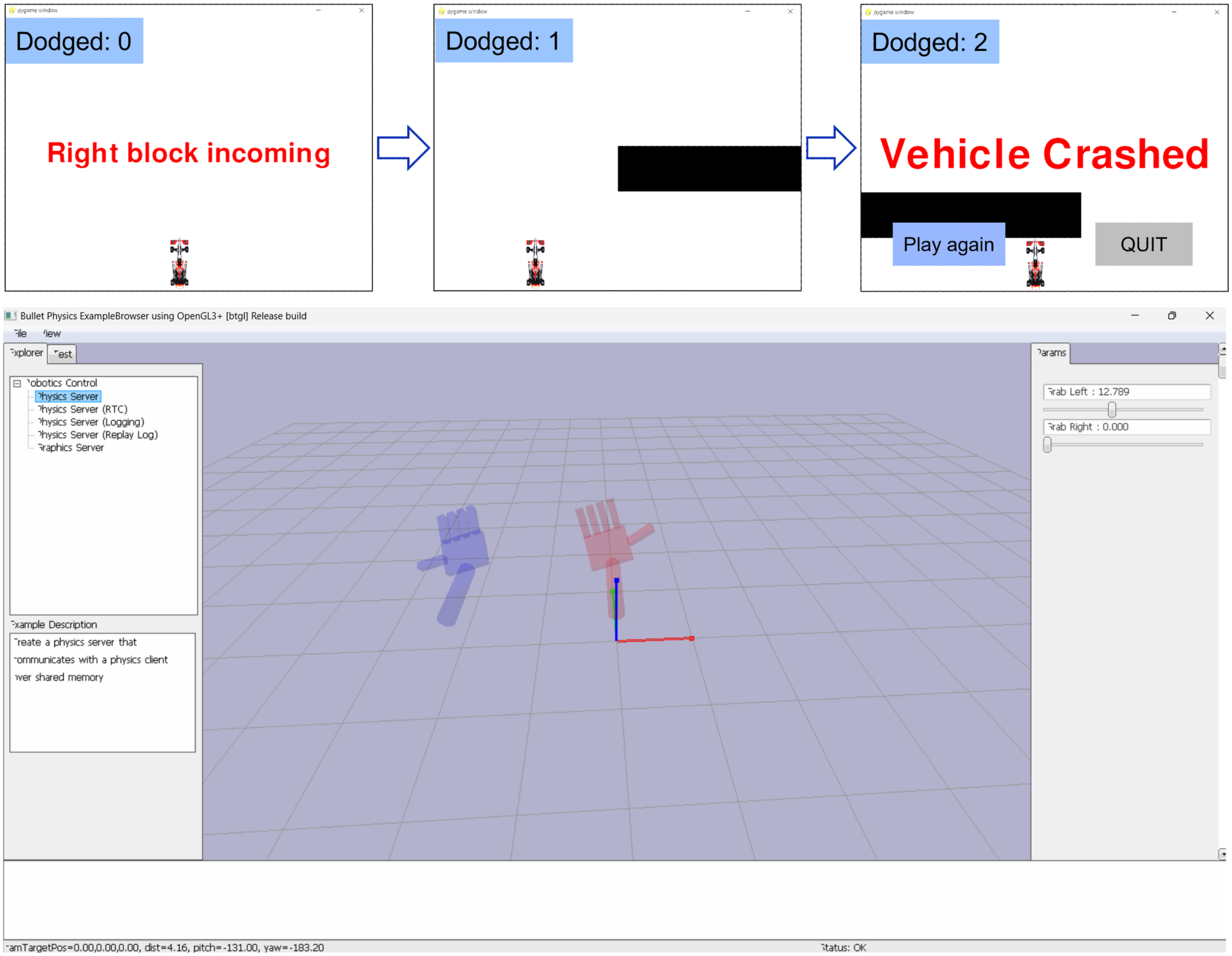

**Fig 9. Illustration of PyNoetic's 2D and 3D simulation module with visual feedback.**

For the 3D simulation, PyNoetic uses the URDF (Unified Robotics Description Format) to create a virtual model of two arms for 02 class BCI tasks. URDF is an XML representation of a robotic multi-body system. PyBullet [58] is used to create a visual representation of the URDF file and interface with other PyNoetic modules. At present, PyNoetic uses the grasping of fingers on one of the hands as a visual indicator of model inference. More degrees of freedom can be added by modifying the URDF file for more than two class classification tasks. The 3D simulation is shown in Fig 9.

**(vii) PyNoetic's visualization module.** Data visualization plays a pivotal role in providing clear insights and validating algorithms employed in BCI research, rather than using them as a black box. Real-time visualization is particularly valuable for detecting outliers, debugging algorithms, and providing immediate feedback on BCI performance. Additionally, visualization aids in implementing artifact removal techniques, such as selecting Independent Component Analysis (ICA) components for artifact elimination from EEG data [53]. PyNoetic leverages PyQTGraph [59], an extension of PyQT [60] to construct an intuitive user-friendly interface for data visualization. Within PyNoetic's visualization module, a versatile array of plots can be generated, including raw EEG activity, smoothed EEG activity, frequency responses of filters, filtered EEG data, ICA components, artifact-free EEG activity, Fourier Transforms of EEG data, and Welch's Periodograms, among others. Further, it supports dynamic, interactive graphs, allowing users to seamlessly pan, scale, and isolate specific data segments effortlessly using a mouse. These visualization capabilities serve as invaluable tools for researchers, providing deep insights into BCI system performance and data characteristics. They enable detailed analysis and refinement of algorithms, fostering advancements in BCI technology and applications.

## 4 Design principles

**Parallel processing.** To optimize the framework's performance, a modular approach with streamlined parallel processing is employed. In the stimuli generation and recording module, data recording operates on a dedicated processor thread separate from the stimuli presentation. This design choice ensures that PyNoetic's data acquisition sub-module captures EEG data autonomously without impacting overall system responsiveness. This data can be acquired through proprietary APIs such as Cortex for Emotiv [61] devices or directly from EEG devices that support the Lab Streaming Layer and its Python implementation, PyLSL, which PyNoetic supports. Data can also be streamed through a custom implementation employed by the user for their hardware. Once the data recording is concluded, the data is transferred to the main thread for processing. Again, the processing thread is separate from the graphical user interface thread. The thread structuring ensures responsiveness for demanding computations like Independent Component Analysis.

**Varied recording hardware compatibility.** While PyNoetic seamlessly integrates with Emotiv devices, it supports effortless compatibility extension with alternative EEG devices through a modular Python script, executed by an autonomous thread. Various Python libraries are readily available to facilitate both wired and wireless communication with third-party sensors, thereby accommodating diverse recording hardware in EEG data acquisition.

**Online mode.** PyNoetic is designed to operate in two distinct modes: offline and online, each featuring its GUI (see Fig 10). The online mode includes a programmable pick-and-place flowchart architecture, enabling users to select and integrate various stages into their pipeline. This mode, reminiscent of LabVIEW, enables customizable pipeline configurations tailored to real-time EEG signal processing. PyNoetic is the first Python framework (the only other framework is OpenVibe, based in C++) to provide a programmable pick-place flowchart-based GUI for developing online BCI systems, simplifying the creation and modification of processing pipelines.

**Interactive GUI.** Offering a high level of flexibility, PyNoetic provides both low and high levels of GUI-powered abstraction. The online mode GUI facilitates real-time EEG signal processing and responsive system activation for online applications. The GUI accompanying the online mode also offers a programmable pick-and-place flowchart, allowing users the flexibility to modify the processing pipeline effortlessly. Users can effortlessly save and reload pipelines with the click of the mouse, streamlining experimentation procedures. The entire pipeline can be managed within a unified single-screen interface.

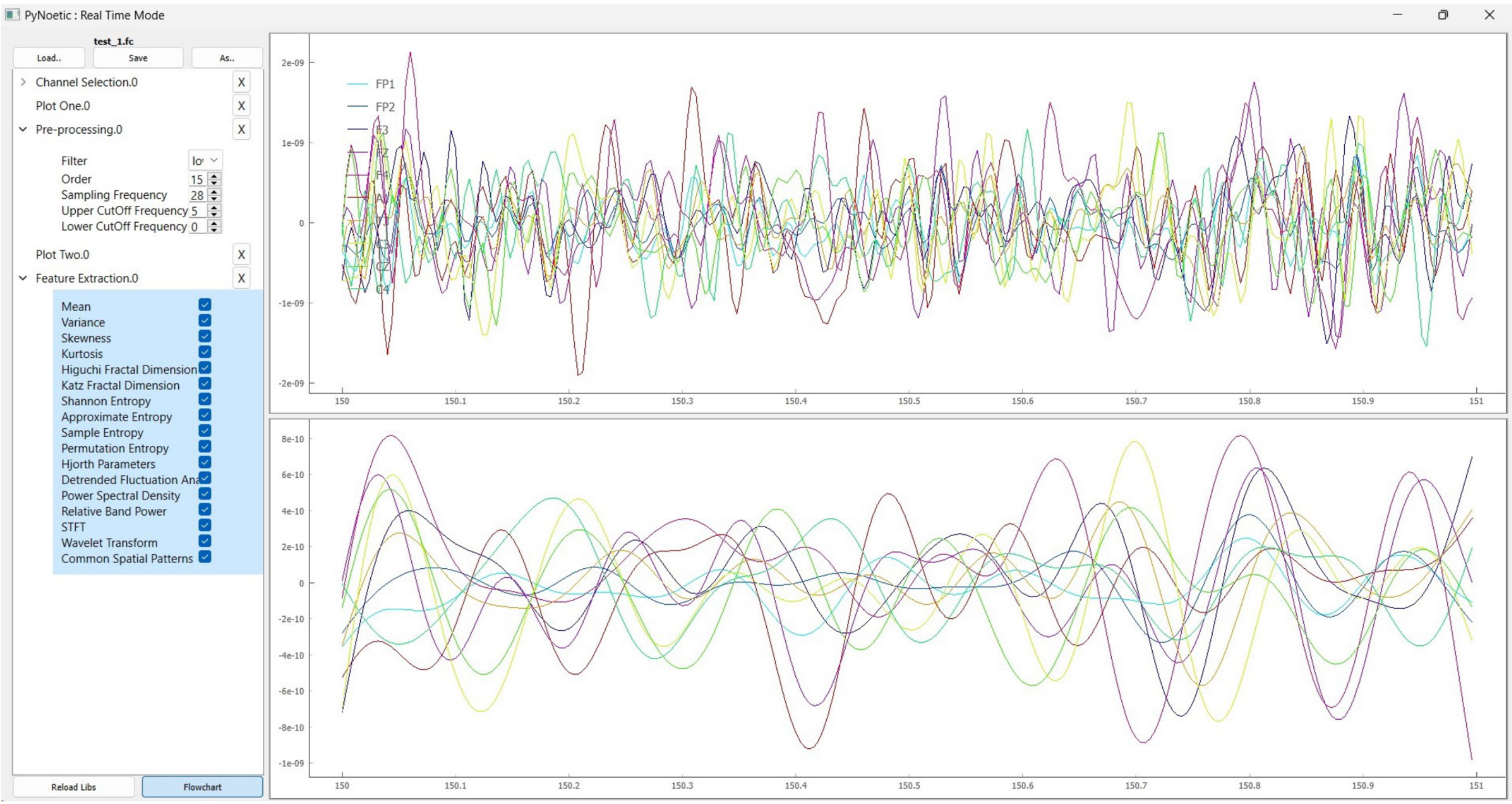

**Fig 10. PyNoetic's online mode in action. Data is streamed from an Emotiv EPOC headset.** The top plot shows the raw data, and the bottom plot shows the filtered EEG data in real time. This is in conjunction with the flowchart shown in Fig 11, where the raw EEG data goes to plot one and the filtered data goes to plot two.

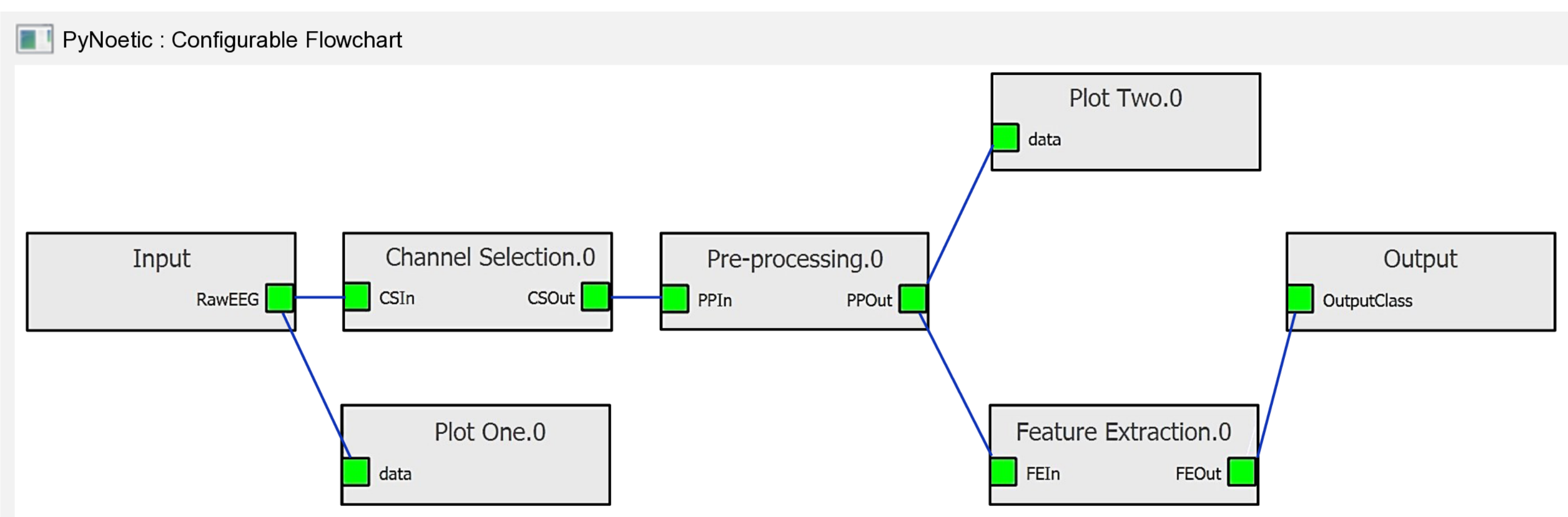

**Fig 11. PyNoetic's unique pick-place configurable flowchart that offers a no-code option for non-programmers.**

**Latency.** While most signal processing algorithms remain consistent across both modes, PyNoetic's online mode incorporates specific optimizations aimed at minimizing latency. PyNoetic is capable of handling offsets in the recorded data with respect to the presented stimuli due to constraints such as network latency by accounting for the mismatch in the timestamps of stimuli presented and data recorded. Another notable optimization is the implementation of amplitude-based faulty epoch rejection, a method designed to effectively eliminate artifacts and reduce computational overhead compared to ICA [61], which, while powerful, can introduce latency [62]. An alternative to ICA is the regression method, which relies on calibration data to remove artifacts; however, this data may not always be available [10]. Certain artifacts, such as those caused by Electrooculography (EOG), often exhibit significantly higher amplitudes than the ongoing background EEG activity [63], a characteristic that can extend to other types of artifacts depending on their nature and electrode placement. Consequently, PyNoetic offers a mechanism to discard flawed epochs based on the amplitude of the acquired signal. It's essential to emphasize that the online and offline modes of PyNoetic are designed to complement each other. The offline mode serves as a platform for advantageous channel selection and classifier training, with the selected channels and trained classifiers subsequently utilized by the online mode for real-time signal classification. This holistic approach ensures efficient integration of offline analysis outcomes into online operations.

## 5 Methods: Participant data collection

The study involves the collection of non-invasive EEG data for software testing. Ten participants were recruited between August 1, 2022, to September 1, 2022. Prior ethical approval was obtained in writing from the Institutional Ethics Committee (IEC) of Thapar Institute of Engineering and Technology, India. Written consent of the participants was obtained before recording the sample EEG dataset. No minors were involved in the study. The data was anonymized to ensure the protection of participants' rights and confidentiality throughout the study.

## 6 Experimental verification and case studies

**Online Analysis.** An Emotiv headset was used to record the participant's EEG data following the SSVEP paradigm. An illustration of the recording paradigm and the corresponding real-time channel selection and pre-processing is shown in Fig 12. The corresponding results of the ICA performed on PyNoetic are shown in Fig 13.

**Motor imagery data decoding experiment.** To test PyNoetic's classification modules, we processed and classified existing datasets, i.e., Motor Imagery dataset [64]. It consists of EEG recordings of 8 participants performing three tasks, i.e., right-hand motor imagination, feet motor imagination, and rest state. The data was collected at a frequency of 512 Hz using 16 wet electrodes. Corresponding to each task, 20 trials were conducted, and each trial lasted for 3 seconds. We utilized three models from PyNoetic's classification module: Shallow-Net, EEG-Net, and Deep-Net. Table 2 presents the performance results from a 5-fold cross-validation for the models used. While EEG-Net achieved the highest accuracy when averaging across different subjects, both Shallow-Net and Deep-Net demonstrated a negligibly higher average score for the Matthews Correlation Coefficient (MCC) compared to EEG-Net.

It must be noted that the primary goal of this study is to prove PyNoetic's efficacy in developing custom BCI pipelines. The results obtained in this study should not be construed as definitive performance benchmarks due to the small number of subjects that took part in the study.

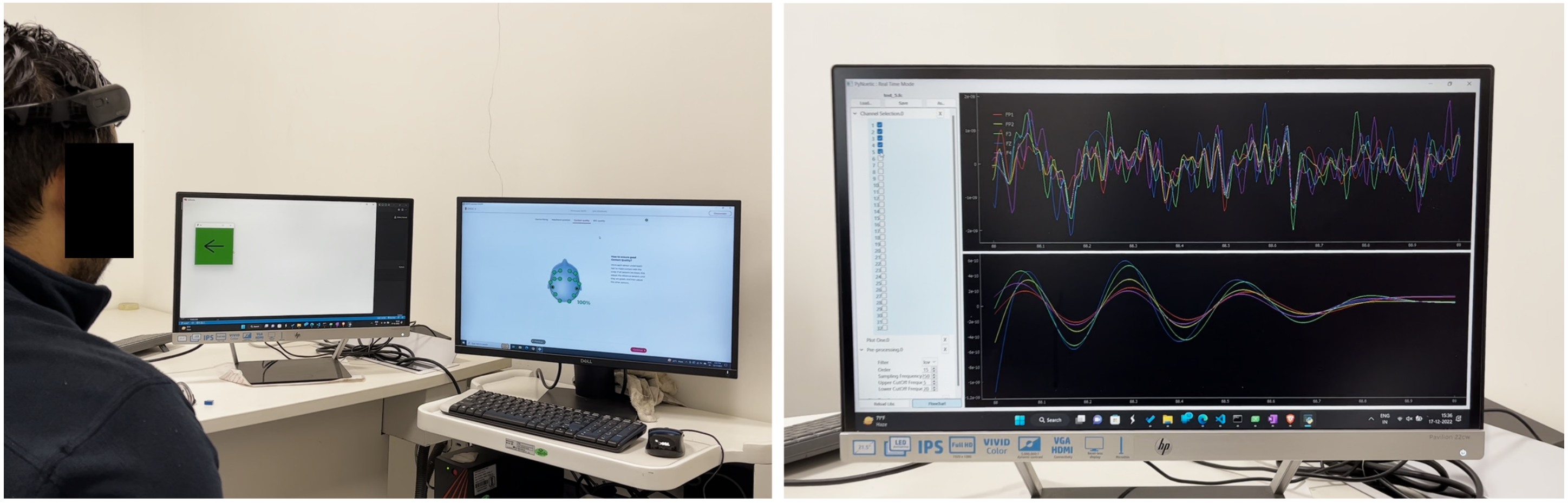

(a) (b)

**Fig 12. Illustration of recording paradigm with PyNoetic's Stimuli generation and recording module.** (a) Picture of an SSVEP recording session. (b) Real-time Channel Selection and preprocessing in online mode.

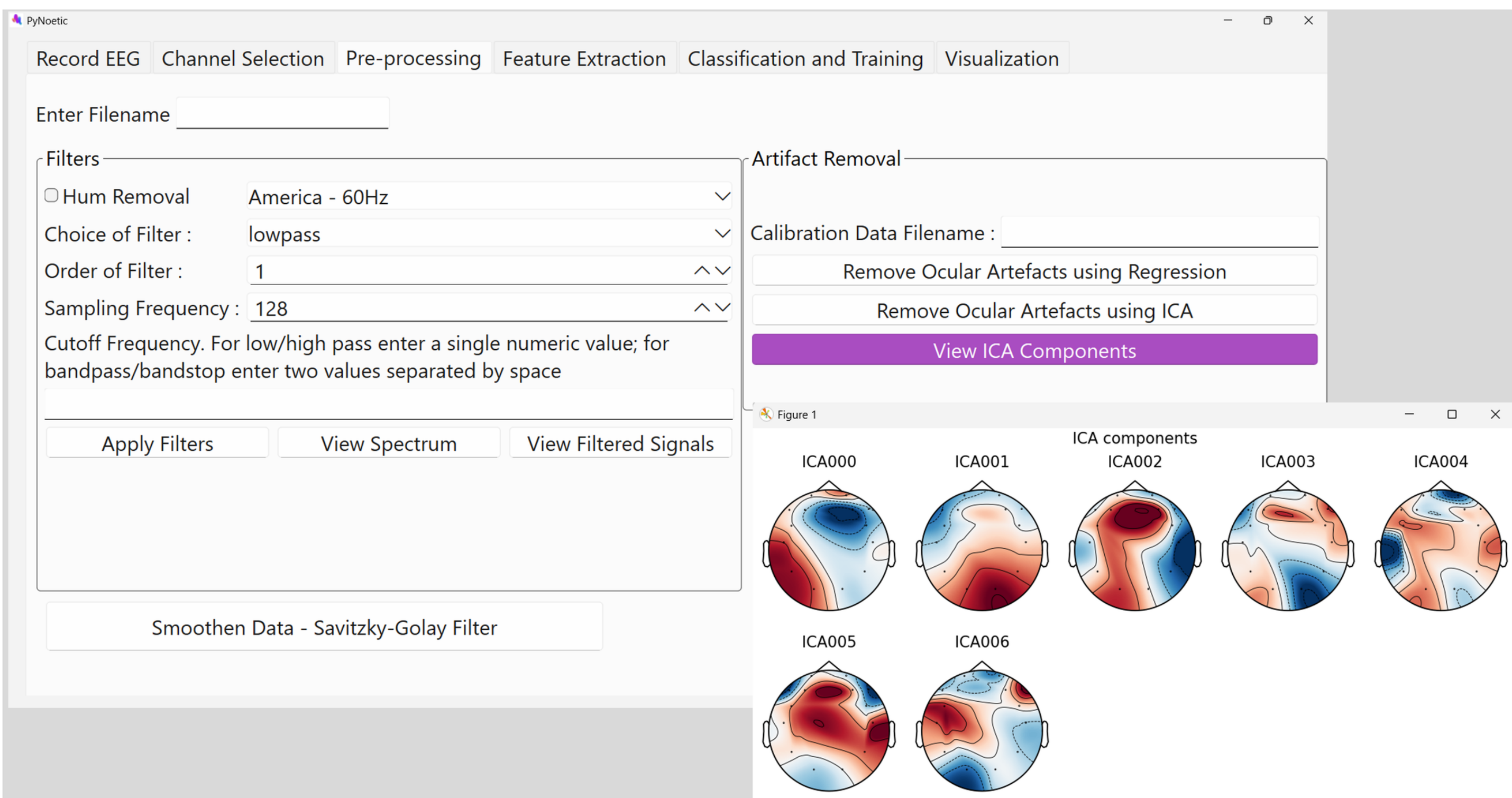

**Fig 13. The results of ICA performed using PyNoetic.**

**Design of oddball and three-oddball paradigms.** The modular architecture of PyNoetic provides extensive flexibility for research, accommodating diverse experimental paradigms within a unified framework. This section details how PyNoetic supports the design and implementation of oddball paradigms. The oddball paradigm, widely employed in neuropsychological experiments, can be effortlessly implemented into PyNoetic's ERP module. It utilizes a dual-stimulus presentation strategy consisting of rare and typical stimuli, with the former appearing less frequently than the latter. Participants are instructed to refrain from responding to typical stimuli but to engage in a task (imaginative or otherwise) upon encountering rare stimuli. Importantly, the oddball paradigm is highly effective in eliciting an ERP response compared to conventional stimulus sequences [65]. PyNoetic enables researchers to customize experimental designs, including the implementation of a three-oddball paradigm. Within the ERP module, users can adjust class distributions to suit their specific requirements. For example, a three-oddball paradigm can be configured by assigning specific weights to each class, such as the distribution 0.1 : 0.1 : 0.8. Here, classes with weights of 0.1 represent rare or "deviant" stimuli, while the class assigned a weight of 0.8 corresponds to typical stimuli. Thus, PyNoetic empowers researchers to effortlessly design and execute both standard and tailored oddball paradigms, significantly enhancing the framework's experimental versatility.

**Table 2. Performance of various deep learning models from PyNoetic's classification module on the Motor Imagery decoding task.**

| Model | Subject-1 | Subject-2 | Subject-3 | Subject-4 | Subject-5 | Subject-6 | Subject-7 | Subject-8 | AVG. |
|---|---|---|---|---|---|---|---|---|---|
| Performance Accuracy (%) | | | | | | | | | |
| Shallow-Net | 82.5 | 70.0 | 92.5 | 80.0 | 65.0 | 77.5 | 67.5 | 70.0 | 75.6 |
| Deep-Net | 75.0 | 80.0 | 60.0 | 77.5 | 65.0 | 75.0 | 65.0 | 75.0 | 71.5 |
| EEG-Net | 77.5 | 75.0 | 90.0 | 87.5 | 70.0 | 80.0 | 75.0 | 82.5 | 79.7 |
| Matthews Correlation Coefficient (MCC) | | | | | | | | | |
| Shallow-Net | 0.85 | 0.31 | 0.61 | 0.90 | 0.74 | 0.75 | 0.65 | 0.74 | 0.69 |
| Deep-Net | 0.75 | 0.58 | 0.64 | 0.90 | 0.52 | 0.87 | 0.72 | 0.46 | 0.68 |
| EEG-Net | 0.71 | 0.41 | 0.61 | 0.85 | 0.55 | 0.67 | 0.54 | 0.76 | 0.64 |

## 7 Discussion and comparison with existing BCI frameworks

Table 3 provides a comparative overview of the functionality across popular Python-based BCI frameworks, including PyNoetic. Fig 12 presents a demo of PyNoetic, and interested readers are encouraged to explore additional demo videos available on the project website.

**End-to-End Support for BCI development.** In a BCI system, each step of the data processing pipeline influences the subsequent steps and the final outcome of experiments. Researchers from different domains working on BCI find themselves writing pieces of SW to process their experimental data, and a direct impact of this is that the quality of research depends on the quality of the framework written [66]. While frameworks like Gumpy [14] offer modular structure and documentation, they lack critical functionalities such as artifact removal and channel selection techniques. Similarly, BioPyC [13], which integrates with Jupyter, lacks real-time capabilities and comprehensive channel selection techniques. Medusa [25] supports a wide variety of BCI paradigms and signal processing methods, including some deep learning methods; however, it lacks real-time charts and support for multi-class motor imagery and is only available on Windows. MetaBCI [67] lacks advanced EEG analysis and decoding methods, such as non-linear feature extraction and connectivity measures. The absence of a GUI makes MetaBCI more suitable for experienced users. Further, BCI-HIL [30] is one of the few BCI software that leverages cloud computing and web-based GUIs to control experiments. The signal processing algorithm is primarily script-driven using TimeFlux [27], but the software has not seen active adoption by the community.

**Table 3. Comparison of PyNoetic's functionality for BCI design with other existing state-of-the-art frameworks.**

| Framework | Publ. | Updated[2] | Stimuli Presentation | Non-linear Features | Algo. Simulation | Offline Analysis | Real-time[3] Development | Connectivity Analysis | Programmable Flowchart | Cross-Platform[4] | Novel Paradigms[5] | GUI | Cites[6] |
|---|---|---|---|---|---|---|---|---|---|---|---|---|---|
| C++/MATLAB Frameworks[1] | | | | | | | | | | | | | |
| BCI2000 [18] | 2004 | 2023 | ✓ | ✗ | ✓ | ✓ | ✓ | ✗ | ✗ | ✓ | ✗ | ✓ | 3402 |
| OpenVibe [19] | 2010 | 2024 | ✓ | ✗ | ✓ | ✓ | ✓ | ✗ | ✓ | ✓ | ✗ | ✓ | 962 |
| BCILAB [71] | 2013 | 2014 | ✓ | ✗ | ✗ | ✓ | ✓ | ✗ | ✗ | ✓ | ✗ | ✓ | 397 |
| EEGLAB [72] | 2004 | 2024 | ✗ | ✗† | ✗ | ✓ | ✗† | ✗ | ✗ | ✓ | ✗ | ✓ | 23107 |
| Python Frameworks | | | | | | | | | | | | | |
| MNE [11] | 2013 | 2024 | ✓ | ✓ | ✗ | ✓ | ✗† | ✓ | ✗ | ✓ | ✓ | ✗† | 2766 |
| Wyrm [12] | 2015 | 2015 | ✗ | ✗ | ✗ | ✓ | ✗ | ✗ | ✗ | ✓ | ✗ | ✗ | 34 |
| BioPyC [13] | 2021 | 2021 | ✗ | ✓ | ✗ | ✓ | ✗ | ✗ | ✗ | ✓ | ✗ | ✓ | 11 |
| Gumpy [14] | 2018 | 2019 | ✓ | ✗ | ✓ | ✓ | ✓ | ✗ | ✗ | ✓ | ✓ | ✗ | 43 |
| PyFF [23] | 2010 | 2011 | ✓ | ✗ | ✗ | ✗ | ✓ | ✗ | ✗ | ✓ | ✗ | ✓ | 80 |
| PyEEG [33] | 2011 | 2021 | ✗ | ✓ | ✗ | ✓ | ✗ | ✗ | ✗ | ✓ | ✗ | ✗ | 195 |
| eeglib [34] | 2021 | 2021 | ✗ | ✓ | ✗ | ✓ | ✗ | ✗ | ✗ | ✓ | ✗ | ✗ | 23 |
| EEGraph [35] | 2023 | 2023 | ✗ | ✓ | ✗ | ✓ | ✗ | ✓ | ✗ | ✓ | ✗ | ✗ | 6 |
| Medusa [25] | 2023 | 2024 | ✓ | ✓ | ✓ | ✓ | ✓ | ✓ | ✗ | ✗ | ✓ | ✓ | 25 |
| BESP [73] | 2023 | 2023 | ✓ | ✗ | ✓ | ✓ | ✗ | ✗ | ✗ | -‡ | ✗ | -‡ | - |
| NeuroKit2 [36] | 2021 | 2024 | ✗ | ✓ | ✗ | ✓ | ✗ | ✓ | ✗ | ✓ | ✗ | ✗ | 781 |
| HappyFeat [26] | 2024 | 2024 | ✗ | ✗ | ✗ | ✓ | ✗ | ✓ | ✓ | - | ✗ | ✓ | - |
| MetaBCI [67] | 2024 | 2024 | ✓ | ✗ | ✗ | ✓ | ✓ | ✗ | ✗ | - | ✓ | ✗ | - |
| BciPy [29] | 2021 | 2025 | ✓ | ✗ | ✗ | ✓ | ✓ | ✗ | ✗ | ✓ | ✓ | ✓ | 25 |
| PyBci [74] | 2023 | 2024 | ✗ | ✓ | ✗ | ✓ | ✓ | ✗ | ✗ | ✓ | ✗ | ✗ | 1 |
| Neuxus [28] | 2022 | 2023 | ✗ | ✗ | ✗ | ✓ | ✓ | ✗ | ✗ | ✓ | ✗ | ✗ | 5 |
| BCI-HIL [30] | 2023 | 2023 | ✓ | ✗ | ✓ | ✓ | ✓ | ✗ | ✗ | ✓ | ✓ | ✓ | 5 |
| BCI Toolbox [75] | 2024 | 2024 | ✗ | ✗ | ✓ | ✓ | ✓ | ✗ | ✗ | ✓ | ✗ | ✓ | 1 |
| PsychoPy [76] | 2007 | 2025 | ✓ | ✗ | ✗ | ✗ | ✓ | ✗ | ✗ | ✓ | ✗ | ✓ | 5656 |
| Time Flux [27] | 2019 | 2024 | ✗ | ✗ | ✗ | ✓ | ✓ | ✗ | ✗ | ✓ | ✓ | ✗ | 14 |
| PyNoetic | - | 2025 | ✓ | ✓ | ✓ | ✓ | ✓ | ✓ | ✓ | ✓ | ✓ | ✓ | - |

[1] Indicates core programming language used in framework development as per the official paper. Recently, many researchers have developed third-party wrappers to integrate the frameworks into another language. OpenVibe/BCI2000 has plugins that allow integration with Python/MATLAB/LUA scripts, but recommends implementing functions in C++ once the prototype is finalized.
[2] Last Update based on Github package "release". In case of no active release, we report the latest repo commit date.
[3] Ability to deploy the entire pipeline on real-time data.
[4] Refers to support for more than one operating system (Windows, Linux, and macOS). All frameworks (except MEDUSA) are cross-platform.
[5] Ability to implement a novel end-to-end BCI pipeline in real-time, incorporating recent BCI paradigms such as automatic IC rejection, ML models, etc.
[6] Google Scholar Citations up to July 2024 to illustrate popularity.
† There exist some third-party plugins that can be used. ‡ Necessary information could not be found.

Addressing these gaps, PyNoetic provides robust support throughout the BCI development process. It allows for custom stimuli generation for recording new datasets, incorporates sophisticated channel selection algorithms, applies effective artifact filtering and pre-processing methods, and offers a wide array of feature extraction techniques (see Figs 14 and 15). The framework integrates popular classifiers for feature classification and includes simulation tools to validate the efficacy of the developed BCI. Compared to other frameworks, PyNoetic is one of the *first* Python-based BCI frameworks that support end-to-end BCI development with such extensive functionality.

**Pick-place flowchart-based no-code framework.** While many BCI frameworks offer scripting capabilities for customization, it poses a challenge for BCI researchers without programming expertise. To cater to researchers without a background in programming, PyNoetic offers: *(i)* a GUI that simplifies parameter selection and updates within the BCI paradigm via a simple mouse click. PyNoetic's GUI makes the BCI development highly simple to understand. Each tab in the GUI corresponds to a step in the BCI pipeline, allowing users to adjust parameters effortlessly with simple mouse clicks, requiring no prior programming knowledge. *(ii)* Secondly, we introduce a programmable pick-place flowchart-based no-code framework (see Fig 11). Furthermore, PyNoetic offers the ability to develop real-time algorithms with a flow chart-like design and a no-code framework, which is missing in all BCI-Python frameworks. This approach enhances accessibility for researchers new to BCI development, while still providing scripting options for advanced users.

**Transition from MATLAB and C++ to python.** The choice of programming language significantly influences BCI framework adoption. While traditional BCI frameworks like xBCI [16], BF+ [17], BCI2000 [18], and OpenVibe [19] are based on C++, they have low abstraction, require more time for development, and are better suited for experienced researchers. Python's popularity has recently surged due to its ease of use and extensive library support (e.g., SciPy [21], Numpy [24], etc.), and reduces development time and cost. Moreover, most ML and deep learning architectures have Python support compared to C++. PyNoetic capitalizes on Python's strengths while addressing potential performance concerns through optimized paradigms like parallel processing and latency reduction (as discussed in the SW design section). Unlike newer frameworks such as Medusa[25], which lacks Linux and macOS support (as per the latest documentation), PyNoetic ensures cross-platform compatibility, accommodating users on Windows, macOS, and Linux environments.

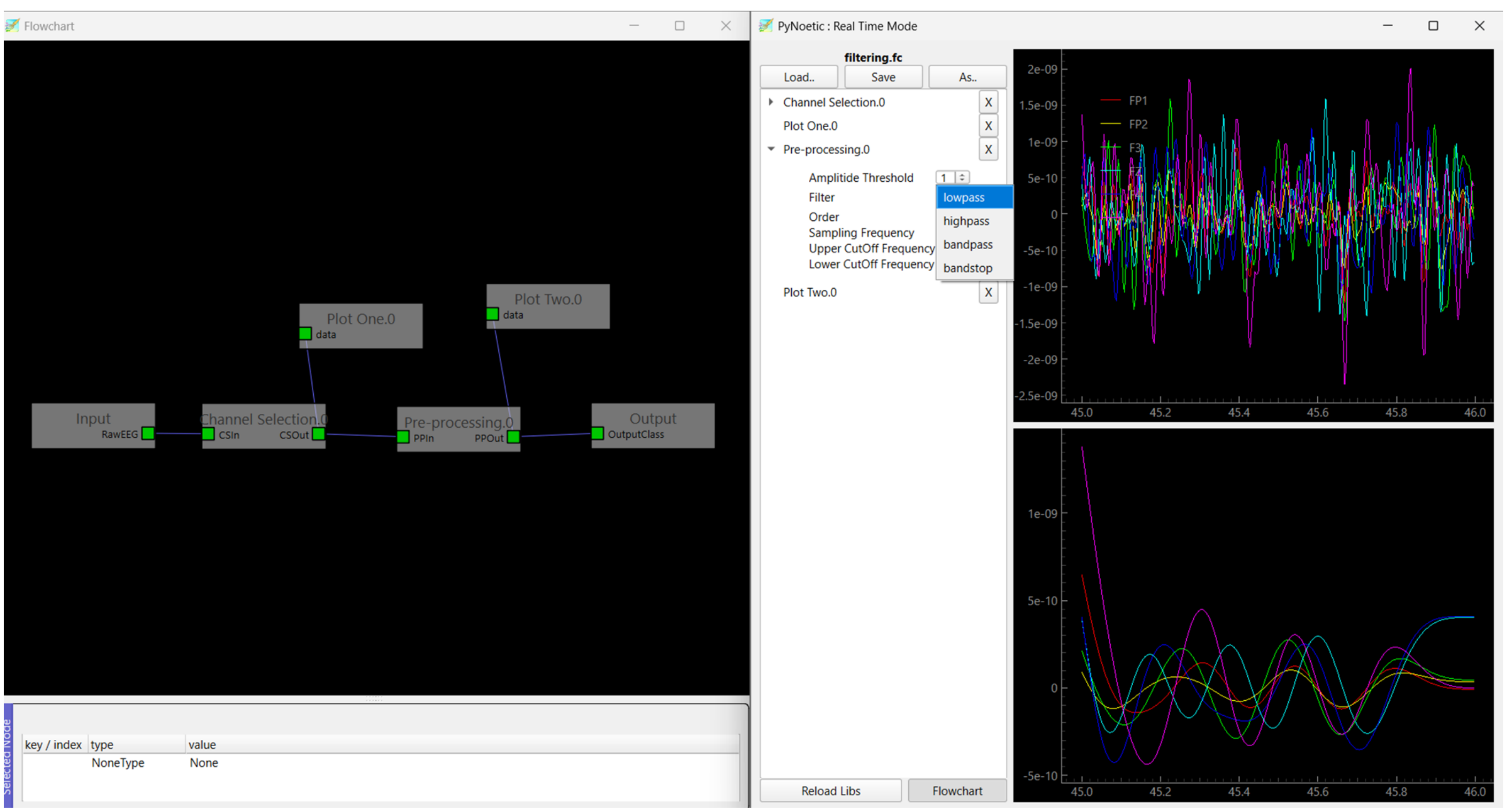

**Fig 14. Pseudo live-stream of EEG data is generated, and a simple pick-and-place flowchart is designed for channel selection and filtering.** The top plot displays the raw EEG signal, while the bottom plot shows the filtered EEG signal, with each instance representing data from a single epoch.

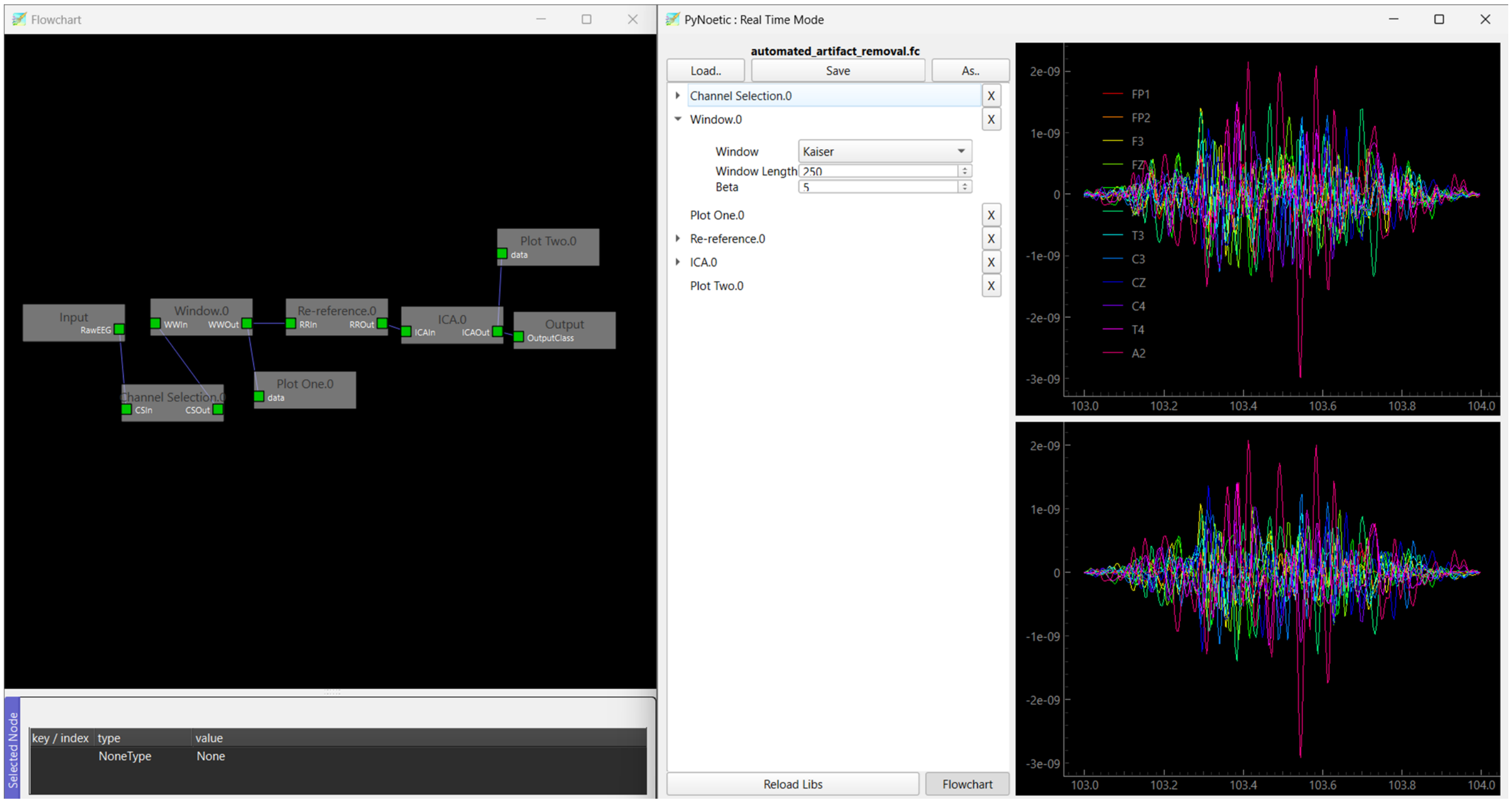

**Fig 15. Pseudo live-stream of EEG data is processed through the flow chart: after channel selection, each epoch undergoes Kaiser windowing (length = 250), followed by re-referencing to the common average.** The ICA block employs MNE-ICA to detect and remove artifacts from the EEG data, reconstructing the signal with the remaining components. The right EEG signal plots display the same before and after the process.

**Modularity to enable community SW updates.** An indispensable aspect of SW frameworks is the need for regular updates to ensure it is current and relevant. A notable Python-based BCI framework, Pyff [23], was discontinued in 2016. Similarly, Wyrm [12], which initially aimed to reduce redundancy in reprogramming standard paradigms and facilitate reproducible research, has seen no active contributions since 2016. With few exceptions, such as the MNE [11], OpenVibe [19], which have undergone substantial updates, most BCI frameworks have stagnated without functional improvements. For instance, BCI2000 [18] has received updates, but has not expanded its functionalities. Other frameworks have either been discontinued or are limited to specific user groups who initiated their development. This highlights the prevailing challenge in the BCI community regarding SW maintenance, which restricts the broader adoption of these frameworks.

A key reason for the lack of active community updates is that BCI development is a multi-faceted task, and demands proficiency across diverse domains such as programming, neuroscience, signal processing, ML, electroencephalography, embedded systems, SW development, and operations. To address this challenge, PyNoetic has been carefully compartmentalized into 7 modules (see Fig 16), each catering to distinct areas of expertise within the BCI research community. This modular design empowers domain experts to update specific aspects of the SW without affecting other components. Moreover, the modular architecture of PyNoetic allows researchers to incorporate features from other Python packages that may be absent in the framework.

## 8 Conclusion

The current study presents PyNoetic, a novel, free, and open-source framework designed for the development, testing, and prototyping of BCIs. Developed in Python, a widely adopted open-source language, PyNoetic offers an alternative to the traditionally dominant tools such as MATLAB and C++. PyNoetic is aimed at offering a platform that simplifies the rapid prototyping and development of BCI systems. It enables users to evaluate algorithms offline and deploy them in real-time scenarios. Encompassing various stages of BCI development, PyNoetic supports functions ranging from stimulus generation and data acquisition to simulation and feedback mechanisms. Notably, it stands out as the first cross-platform Python

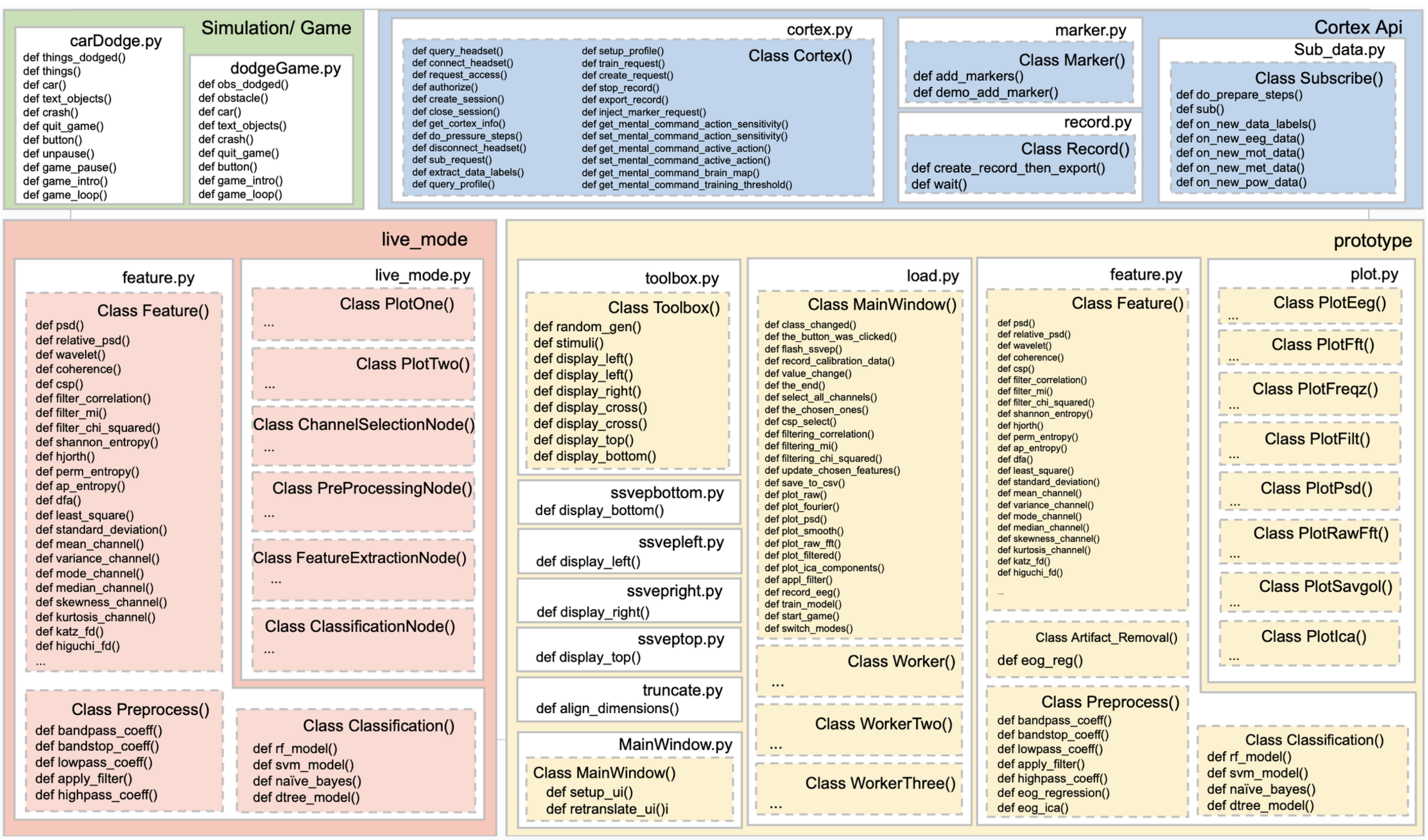

**Fig 16. Modular architecture design of PyNoetic showing all its constituent functions.**

https://doi.org/10.1371/journal.pone.0327791.g016

framework to provide a GUI for real-time application development in the BCI domain. These capabilities position PyNoetic as one of the most comprehensive BCI platforms available today. We anticipate that PyNoetic will garner acceptance and contributions from the open-source community, evolving beyond its current state. While PyNoetic primarily focuses on synchronous BCI, our vision extends towards enabling seamless interaction with surroundings through BCI, reducing reliance on visual or auditory stimuli. While currently emphasizing synchronous BCI paradigms, our vision extends towards enabling seamless interaction with surroundings through BCI, reducing reliance on visual or auditory stimuli. Moreover, the framework's flexible, modular, and scalable design facilitates ongoing maintenance and continuous development to meet evolving BCI research and application needs.

## 9 Limitations and future work

Subsequent developments in PyNoetic are focused on expanding the framework's functionality, including enhancing interoperability between different file formats, introducing additional features, and refining the GUI. Notably, existing Python frameworks with GUIs often lack implementations of Channel Selection algorithms, a gap that PyNoetic addresses by integrating channel selection algorithms based on filtering techniques [68]. Although these techniques offer speed and scalability, they may compromise accuracy by not considering combinations of multiple channels, as highlighted in previous literature [42]. Other channel selection methods based on wrapper and embedded techniques can also be included in the framework to broaden its capabilities. Moreover, PyNoetic currently supports artifact removal using regression and ICA. However, it is acknowledged that both techniques have limitations, which have been addressed by newer implementations [69,70]. To enhance artifact removal capabilities, PyNoetic aims to integrate these advancements into its framework. While BCI paradigms like Motor Imagery and SSVEP are a part of PyNoetic, the stimuli presentation and data recording modules could be enhanced and improved by the incorporation of methods for drift correction during longer recordings that are currently missing. As the toolbox garners acceptance and contributions from the open-source community, more such limitations shall be identified and overcome.

## Author contributions

**Conceptualization:** Gursimran Singh, Rahul Upadhyay.

**Data curation:** Gursimran Singh, Rahul Upadhyay.

**Formal analysis:** Gursimran Singh.

**Investigation:** Gursimran Singh, Rahul Upadhyay.

**Methodology:** Gursimran Singh, Rahul Upadhyay.

**Project administration:** Rahul Upadhyay.

**Resources:** Rahul Upadhyay.

**Software:** Gursimran Singh.

**Supervision:** Rahul Upadhyay, Vinay Kumar, Luca Longo.

**Validation:** Rahul Upadhyay.

**Visualization:** Aviral Chharia, Vinay Kumar, Luca Longo.

**Writing – original draft:** Gursimran Singh.

**Writing – review & editing:** Aviral Chharia, Rahul Upadhyay, Vinay Kumar, Luca Longo.